\documentclass[aps,prd,showpacs,preprintnumbers,twelvepoint,floatfix]{revtex4}
\usepackage{epsfig}
\usepackage{latexsym,amsmath,amssymb,amstext}
\usepackage{float}
\usepackage{graphicx}
\usepackage{color}

\newcommand\bef{\begin{figure}}
\newcommand\eef[1]{\label{fg:#1}\end{figure}}
\newcommand\beq{\begin{equation}}
\newcommand\eeq[1]{\label{#1}\end{equation}}
\newcommand\beqa{\begin{eqnarray}}
\newcommand\eeqa[1]{\label{#1}\end{eqnarray}}
\newcommand\bet{\begin{table}}
\newcommand\eet[1]{\label{tb:#1}\end{table}}

\newcommand\fgn[1]{Figure \ref{fg:#1}}
\newcommand\eqn[1]{eq.\ (\ref{#1})}

\newcommand\apx[1]{Appendix \ref{sec:#1}}
\newcommand\tbn[1]{Table \ref{tb:#1}}

\newcommand\ie{{\sl i.e.\/}}

\newcommand\etal{{\sl et al.\/}}
\newcommand\jhep{{\sl J.\ H.\ E.\ P.\/}\ }
\newcommand\np{{\sl Nucl.\ Phys.\/}\ }

\newcommand\pos{{\sl PoS\/}\ }

\newcommand\plt{{\sl Phys.\ Lett.\/}\ }

\newcommand\op[1]{{\mathcal O}_{#1}}
\newcommand\ppb{\ensuremath{\langle\overline\psi\psi\rangle}}

\newcommand{\mub}{\mu_{\scriptscriptstyle B}}
\newcommand{\mube}{\mu_{\scriptscriptstyle B}^{\scriptscriptstyle E}}

\begin{document}
\title{The equation of state of QCD at finite chemical potential}
\author{Sourendu\ \surname{Gupta}}
\email{sgupta@theory.tifr.res.in}
\affiliation{Department of Theoretical Physics, Tata Institute of Fundamental
         Research,\\ Homi Bhabha Road, Mumbai 400005, India.}
\author{Nikhil\ \surname{Karthik}}
\email{nikhil@theory.tifr.res.in}
\affiliation{Department of Theoretical Physics, Tata Institute of Fundamental
         Research,\\ Homi Bhabha Road, Mumbai 400005, India.}
\author{Pushan\ \surname{Majumdar}}
\email{tppm@iacs.res.in}
\affiliation{Department of Theoretical Physics, Indian Association for the
         Cultivation of Science,\\ Raja Subodh Chandra Mallick Road,
         Jadavpur, Kolkata 700032, India.}

\begin{abstract}
We obtain the baryon number density, $n$, and the excess contribution
to the pressure, $\Delta P$, at finite chemical potential, $\mub$,
and temperature, $T$, by resumming the Taylor series expansion in a
lattice computation with lattice spacing of $1/(4T)$ and two flavours of
quarks at three different quark masses.  The method proceeds by giving a
critical $\mub$ and limits on the critical exponent, and permits reliable
estimations of the errors in resummed quantities. We find that $n$ and
$\Delta P$ are insensitive to the quark mass.  We also report the bulk
isothermal compressibility, $\kappa$, over a range of $T$ and $\mub$.
\end{abstract}

\preprint{TIFR/TH/14-10}
\maketitle

\section{Introduction}\label{sec:intro}
Heavy-ion collisions produce hot QCD matter which comes to thermal
equilibrium \cite{white}. The resulting fireball evolves. Its evolution is
expected to be described with reasonable accuracy within hydrodynamics
\cite{hydro}.  As a result, computing the equation of state of QCD
matter at values of $T$ and $\mub$ accessible to colliders is a matter
of interest.

A systematic expansion of the pressure, $P$, of QCD as a series in
$\mub$ was first introduced in order to examine the equation of state
at finite chemical potential \cite{mumbai}. The Taylor coefficients of
the expansion are the quark number susceptibilities (QNS), the first of
which had been introduced and studied long ago \cite{milc}. The QNS are
of interest in experimental studies of event-to-event fluctuations of
conserved quantities \cite{ebye,plb}, and therefore have become important
objects of study in recent years. They also indicate a divergence of the
series at finite and real $\mub$, implying the existence of a critical
point \cite{nt4}.

The presence of such divergences could be a barrier to extracting the
equation of state of QCD. In this paper we examine the summation of
these series. We examine techniques of propagating measurement errors and
evaluate the change in pressure due to the baryon chemical potential,
\beq
   \Delta P(T,\mub)= P(T,\mub)-P(T,\mub=0),
\eeq{dpress}
and the baryon number density, $n(T,\mub)$.  Since the pressure at zero
chemical potential, $P(T,\mub=0)$ is being studied with great precision
\cite{hot,wup}, our work opens the door to the evaluation of pressure
over a large part of the phase diagram of relevance to current and
near-future experiments. Interestingly, the method can also begin to
give more information on the critical behaviour \cite{ising}.

This study has another interesting ramification. $\Delta
P$ and $n$ get contributions only from degrees of freedom which
carry baryon number--- baryons at low temperature, and quarks at high
temperature. However, chiral degrees of freedom may be involved through
a mixing between the chiral condensate and the baryon number density
\cite{dynamics}. If this happens, then the scaling directions at the
critical point become mixtures of the physical parameters $T$, $\mub$ and
$m$ (we discuss this point further in \apx{widom}).  In fact, the older
literature sometimes discusses the critical point of QCD purely in terms
of the chiral order parameter. So it is interesting to check how strongly
$\Delta P$ and $n$ depend on the quark mass.  By performing simulations
with three different bare quark masses, we are able to throw some light
on these questions. We study the effect of changing quark masses on the
position of the critical end point and the equation of state, and find
statistically insignificant changes.

Here is the plan of this paper. In the next section we discuss the
simulations and also introduce our notation. The following section is a
detailed technical discussion of all our results, a summary of which also
appears in the final section. The appendices contain technical details
of the analysis of errors on Pad\'e approximants, Widom scaling at the
QCD critical point, and the definition of the isothermal compressibility
in QCD.

\section{Simulations and notation}\label{sec:method}
\bet
\begin{tabular}{|l|c||c|r||l|l|l|l|}
\hline
$\beta$ & $T/T_c$ & Statistics & $\tau$ & $P_s$ & $P_t$ & $W$ & \ppb \\
\hline
5.34 & $0.89\pm0.02$ & 500+10000 & 15 & 1.495(3) & 1.494(4) & 0.041(5) & 0.901(4) \\
5.35 & $0.92\pm0.02$ & 500+20000 & 50 & 1.509(9) & 1.507(7) & 0.05(2) & 0.87(2) \\
5.355 & $0.94\pm0.01$ & 500+20000 & 48 & 1.515(7) & 1.513(6) & 0.06(2) & 0.85(2) \\
5.36 & $0.96\pm0.01$ & 500+40000 & 74 & 1.52(1) & 1.519(8) & 0.06(3) & 0.84(3) \\
5.372 & 1.00 & 500+120000 & 164 & 1.550(7) & 1.545(6) & 0.10(1) & 0.76(2) \\
5.39 & $1.06\pm0.02$ & 500+5000 & 125 & 1.58(1) & 1.58(1) & 0.15(3) & 0.66(5) \\
5.40 & $1.11\pm0.01$ & 500+5000 & 30 & 1.596(3) & 1.587(2) & 0.165(5) & 0.626(9) \\
\hline
\end{tabular}
\caption{The set A runs on $4\times16^3$ lattices and fixed $am=0.1$. 
 Four quantities were monitored to determine thermalization and
 autocorrelations, namely, the spatial plaquette, $P_s$, the temporal
 plaquette, $P_t$, the bare Wilson line, $W$, and the bare quark
 condensate, \ppb. The autocorrelation time, $\tau$, quoted here is the
 maximum of the autocorrelations of these four quantities.}
\eet{runb}

This study was made on $4\times16^3$ lattices using three sets of
configurations, one set each for a different value of $m_\pi/m_\rho$,
as described next. At zero chemical potential, QCD with massive quarks
does not have a phase transition, but a broad crossover. There can be many
conventions for defining the crossover temperature \cite{tc}. We choose
to define it by the peak in the Polyakov loop susceptibility. The peak
in the 4th order QNS occurs at the same temperature \cite{nikhil}.

The first set of configurations (called set A in this paper) was
obtained at $m_\pi/m_\rho\simeq0.58$. Each trajectory was of 1 MD time
unit, and used a time step of $0.05$ MD time units. After discarding
the first 500 trajectories for thermalization, 200 configurations were
collected in each simulation. The details are given in \tbn{runb}. We
identified the critical coupling by a multi-histogram reweighting
and obtained $\beta_c=5.3720\pm0.0005$. This is in agreement with an
old MILC result of $\beta_c=5.375\pm0.020$, obtained on $4\times8^3$
lattice \cite{milc}. At these bare parameters, MILC had made hadron mass
measurements which yielded the value of $m_\pi/m_\rho$ quoted above
\cite{oldmilcm}. Although we do not tune this set to remain at fixed
$m_\pi/m_\rho$, the variation of $\beta$ is small enough that the change
in this ratio over the whole range is expected to be less than 10\%.
We measured the temperature scale given in \tbn{runb} using the method
adopted in \cite{nt4}.

\bet
\begin{tabular}{|ll|c||c|r||l|l|l|l|}
\hline
$\beta$ & $ma$ & $T/T_c$ & Statistics & $\tau$ & $P_s$ & $P_t$ & $W$ & \ppb \\
\hline
5.20 & 0.033 & $0.75\pm0.02$ & 980+4000 & 9 & 1.394(1) & 1.393(1) & 0.01953(3) & 0.9392(1) \\
5.22 & 0.03125 & $0.80\pm0.02$ & 980+4000 & 12 & 1.411(1) & 1.410(1) & 0.02238(3) & 0.9023(7) \\
5.24 & 0.0298 & $0.85\pm0.01$ & 980+4000 & 14 & 1.430(1) & 1.429(1) & 0.02707(7) & 0.856(1) \\
5.26 & 0.02778 & $0.90\pm0.01$ & 980+4000 & 20 & 1.452(2) & 1.450(2) & 0.0351(1) & 0.795(2) \\
5.275 & 0.02631 & $0.95\pm0.01$ & 980+4000 & 30 & 1.472(3) & 1.470(3) & 0.0494(4) & 0.7264(9) \\
5.2875 & 0.025 & 1.00 & 1980+8000 & 98 & 1.503(9) & 1.499(8) & 0.08(1) & 0.60(4) \\
5.30 & 0.02380 & $1.05\pm0.02$ & 980+4000 & 32 & 1.553(3) & 1.543(3) & 0.149(4) & 0.33(2) \\
5.35 & 0.02 & $1.25\pm0.01$ & 980+4000 & 6 & 1.595(1) & 1.584(1) & 0.1858(6) & 0.173(1) \\
5.425 & 0.01667 & $1.50\pm0.01$ & 980+4000 & 6 & 1.6350(4) & 1.6240(3) & 0.2109(4) & 0.1093(2)\\
5.54 & 0.0125 & $2.04\pm0.02$ & 980+4000 & 5 & 1.6830(3) & 1.6730(3) & 0.2393(3) & 0.06630(1) \\
\hline
\end{tabular}
\caption{The set B runs on $4\times16^3$ lattices. 
 Four quantities were monitored to determine thermalization and
 autocorrelations, namely, the spatial plaquette, $P_s$, the temporal
 plaquette, $P_t$, the bare Wilson line, $W$, and the bare quark condensate,
 \ppb. The autocorrelation time, $\tau$, quoted here is the maximum of the
 autocorrelations of these four quantities.}
\eet{runs}

\bef
\includegraphics[scale=0.65]{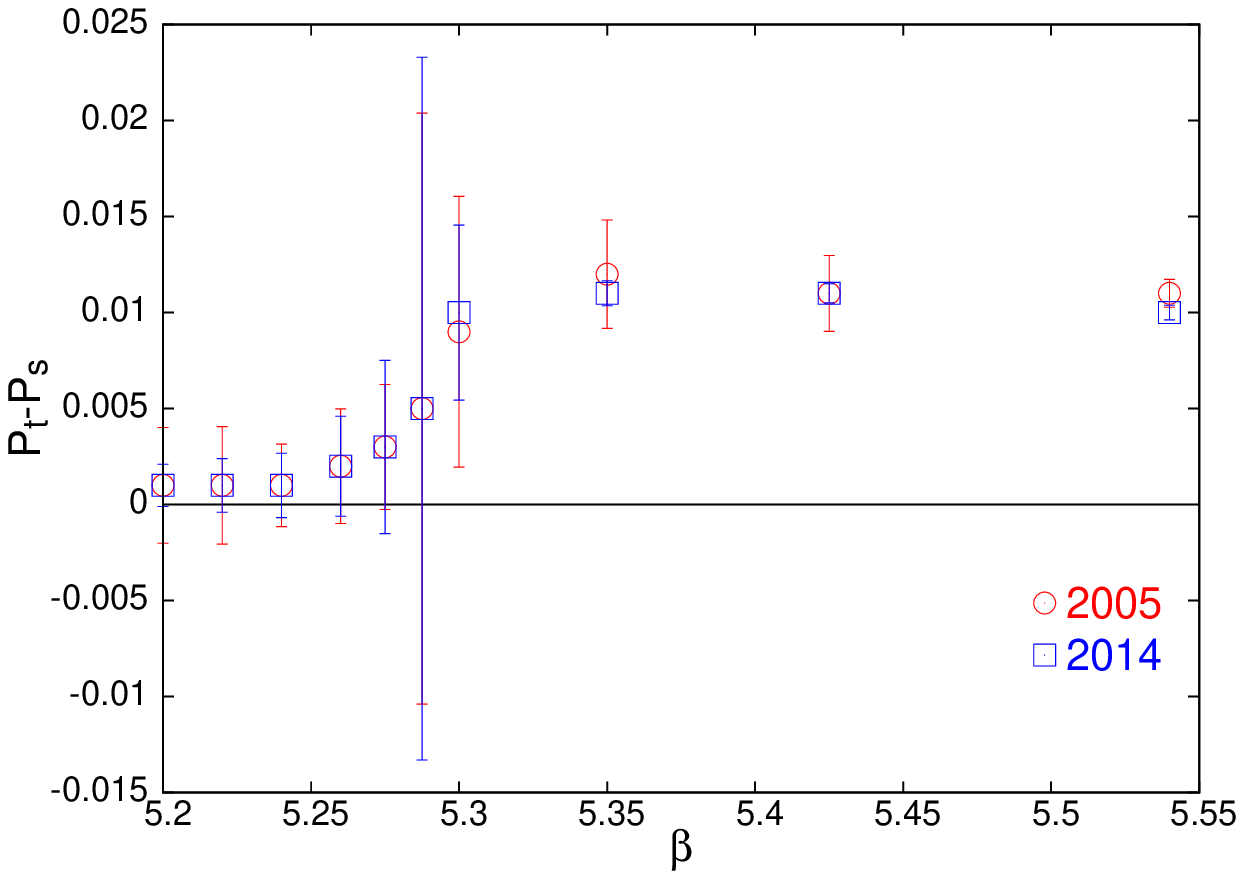}
\includegraphics[scale=0.65]{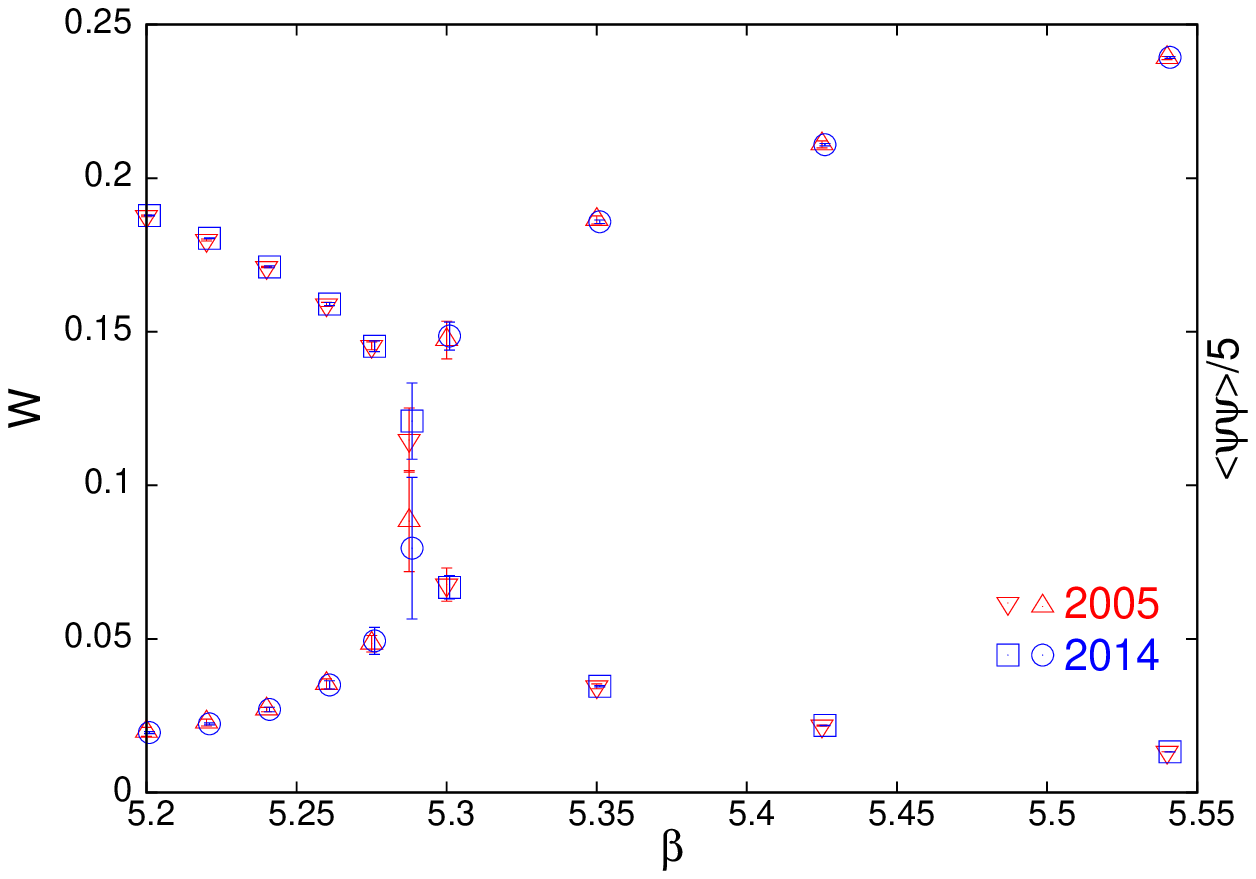}
\caption{Basic comparison of the present set B and older \cite{nt4} runs is
 made through three quantities which enter thermodynamic variables. The first
 panel displays the plaquette difference $P_t-P_s$. The second panel shows
 $W$ (up triangles and circles) and \ppb (down triangles and boxes). The
 large errors at the cross-over coupling, $\beta=5.2875$, are due to
 larger autocorrelations. At all couplings the new simulations are
 compatible with the old, while having smaller errors.}
\eef{comp}

The most extensive set of runs we made was meant to extend the study of
\cite{nt4} with more statistics. The run parameters and statistics of
this set (set B) are given in \tbn{runs}. In these simulations we have
$m_\pi/m_\rho\simeq0.31$.  Each trajectory was of length 2 MD time units,
and used a time step of $0.01$ MD time units.  In each of the runs 980
trajectories were discarded for thermalization.  At $\beta_c=5.2875$ we
collected 100 gauge configurations, one every 80 trajectories, and 200
configurations, one every 20 trajectories, at other values of $\beta$.  This was
sufficient for the stored configurations to be statistically independent.
We checked that the global thermodynamic variables in this run are
fully consistent with those in \cite{nt4}.  These checks are shown
in \fgn{comp}.  The scale setting has been discussed earlier \cite{nt4},
and we use those results to write a temperature scale corresponding to
the bare parameters at which we simulated.

The third set of configurations we used has been described in \cite{nikhil},
where it was called set N. In this paper we call this the set C. For this
set $m_\pi/m_\rho\simeq0.25$ \cite{nikhil}. With about 50--60
configurations, this has lower statistics than the other sets. However,
we used up to 2000 source vectors at several couplings. In order to
understand the effects of statistics, we supplemented it by a sample
of 400 independent configurations at $\beta=5.255$ and $am=0.0165$,
corresponding to $T/T_c=0.93\pm0.01$.  The mass measurements as well as
the determination of the temperature scale was reported in \cite{nikhil}.

\bef
\includegraphics[scale=1.0]{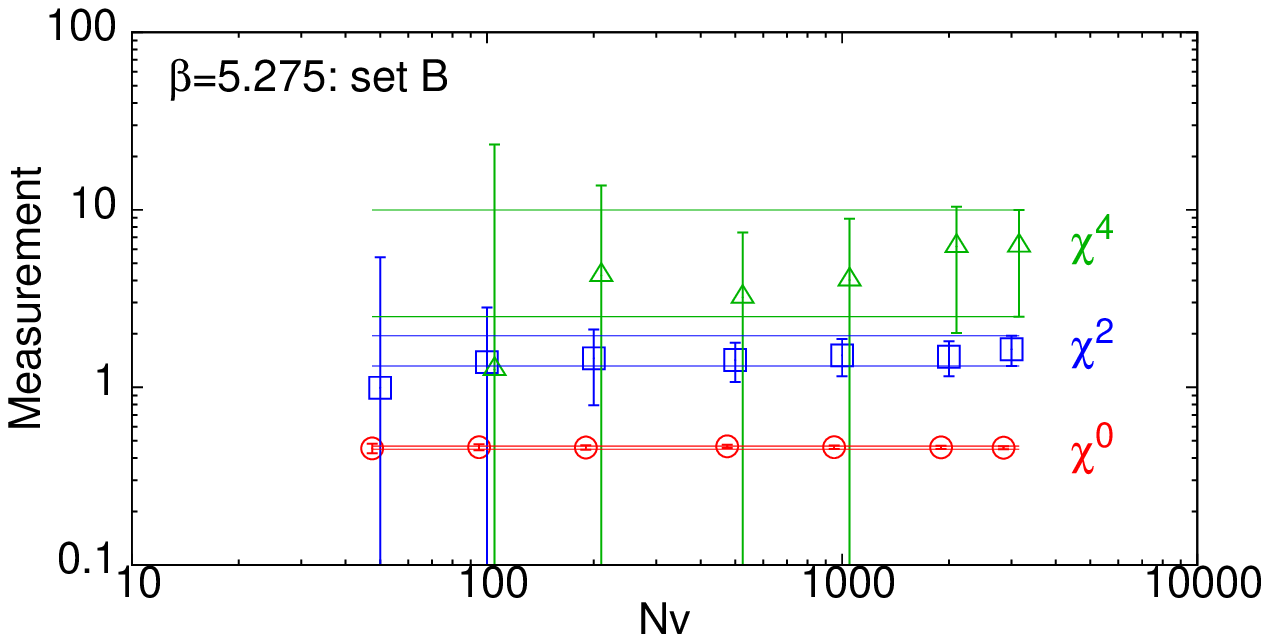}
\caption{The number of fermion sources required to control errors in
 the QNS grows rapidly with the order. The data for $\chi^0$ has been
 displaced to the left by 5\% and that for $\chi^4$ to the right by
 5\%. For the second order QNS, $\chi^0$, it is sufficient to take 50
 sources, whereas for the 6th order QNS, $\chi^4$, one needs at least
 2000 source vectors.}
\eef{stats}

Another difference between this and earlier measurements \cite{nt4}
is the number
of fermion sources, $N_v$, used to perform traces over fermion loops. The
$n$-th order QNS contain up to $n$ loops. We show in \fgn{stats} that
$N_v=50$ seems to suffice for accurate measurements of second order QNS,
but the sixth order requires $N_v=2000$. At $\beta=5.275$ we checked using
$N_v=3000$ that the difference between using $N_v=2000$ and this larger
number of sources is statistically insignificant. So in the rest of the
measurements below $T_c$ we used $N_v=2000$. Above $T_c$ the measurements
turn out to be much less noisy, and we used $N_v=1000$, although it seems
that about half this number of sources is sufficient for our purposes.
The importance of using large $N_v$ was first pointed out in \cite{lat13}.

In \fgn{compqns} we compare the earlier \cite{nt4} and current results.
Clearly the increased statistics leads to reduced errors, as expected,
and also a much smoother variation of results with $T$. The most noticeable
change is that the height of the peak in $\chi_{40}$ is almost half of
its earlier value. By varying $N_v$ and the number of configurations used
in our analysis, we found that this was caused by low statistics.

Many of the measurements we perform involve variables whose distributions
are non-Gaussian, and sometimes skewed \cite{nt6}. All the statistical
analysis in this paper therefore uses the bootstrap procedure,
which yields non-parametric estimators of means and errors. When
the distribution is strongly skewed, the error estimators may be
non-symmetric; in such cases we quote different upper and lower error
bars.

The notation we use in this paper is collected here. The derivatives of the
excess pressure give the baryon number density and the baryon number
susceptibility, respectively,
\beq
   n(T,\mub) = \frac{\partial\Delta P(T,\mub)}{\partial\mub}
      \qquad{\rm and}\qquad
  \chi_B(T,\mub) = \frac{\partial^2\Delta P(T,\mub)}{\partial\mub^2}.
\eeq{defs}
The Taylor expansion of $\Delta P$ at vanishing $\mub$ is
\beq
   \frac{\Delta P(T,\mub)}{T^4} =
         \frac{\chi_B^2(T)}{T^2}\,\frac{\mub^2}{2!T^2}
       + \chi_B^4(T)\,\frac{\mub^4}{4!T^4}
       + \chi_B^6(T)T^2\,\frac{\mub^6}{6!T^6} + \cdots.
\eeq{tse}
The Taylor coefficients are the QNS evaluated at $\mub=0$; only the even
coefficients survive at this symmetric point.
We will
often write $\chi_B$ instead of $\chi_B^2$; no confusion is caused by
this. This Taylor series gives rise to those for $n$ and $\chi_B$.
We work with
two flavours of light degenerate dynamical quarks, so there are in fact
two different conserved quark numbers, one for each flavour. One can
write a double Taylor series expansion
\beq
   \frac{\Delta P(T,\mub)}{T^4} = \sum_{mn}
         \frac1{m!n!}\,\chi_{mn}(T)\,T^{m+n-4}\,\left(\frac{\mu_u}T\right)^m\,
           \left(\frac{\mu_d}T\right)^n
\eeq{taylor}
which defines the usual form of the QNS \cite{mumbai,milc}. Since the
flavours are degenerate, the order of $m$ and $n$ does not matter; the
coefficients with one of them vanishing are called the diagonal QNS.
In the past we have constructed the Taylor expansion 
\beq
   \frac{\chi_{20}(T,\mub)}{T^2} =
         \frac{\chi^0(T)}{T^2}
       + \chi^2(T)\,\frac{\mub^2}{2!T^2}
       + \chi^4(T)T^2\,\frac{\mub^4}{4!T^4} + \cdots
\eeq{div}
and examined its divergence to obtain estimates of the critical end point
\cite{nt4,nt6,lat13}. Here we cross check this result using the Taylor
expansion of \eqn{tse}.

It is useful to recall that the various QNS at any given order have
contributions from a variety of operator topologies \cite{mumbai}, which
can be constructed as follows. In the continuum theory, the operators
contributing to any $N$-th order QNS have $N$ insertions of $\gamma_0$
connected in all possible ways by fermion propagators. The topologies
track the number of insertions within each closed fermion loop; for
example ${\cal O}_{112}$ is a topology which contributes to $N=4$ and
contains two fermion loops with one insertion each and one fermion loop
with two insertions. These topologies also classify the lattice operators.
As a result, any QNS can be written in the form
\beq
   \chi_i = \sum_{\alpha=1}^M n_{i\alpha}\langle{\cal O}_\alpha\rangle,
\eeq{topol}
where $i$ is a labelling of the QNS, $M$ is the number of distinct topologies
of a QNS of that order, $\alpha$ labels the topology, the expectation value
denotes a connected part averaged over configurations, and an enumeration of
$n_{i\alpha}$ is given in \cite{nt4}. This decomposition will turn out to be
useful in analyzing errors.

\bef
\includegraphics[scale=0.7]{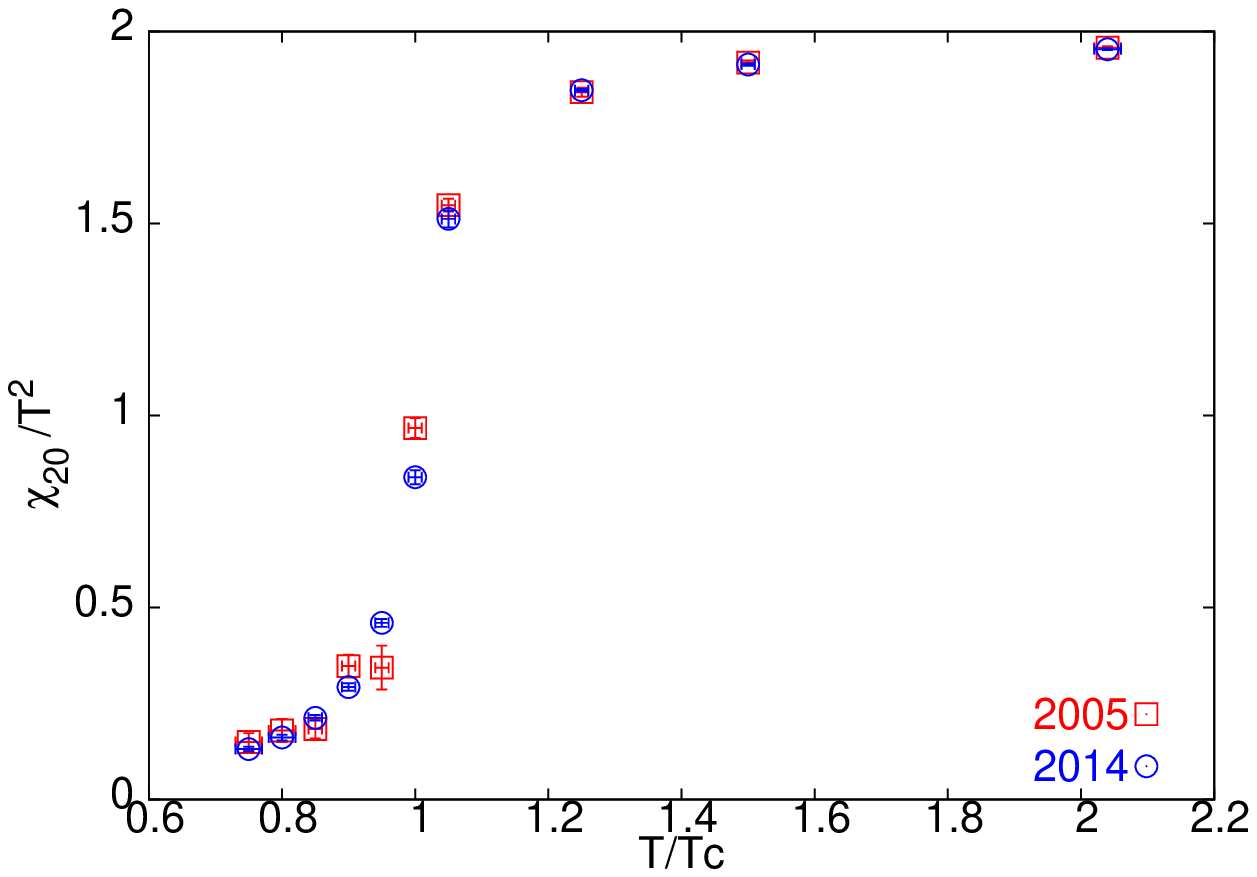}
\includegraphics[scale=0.7]{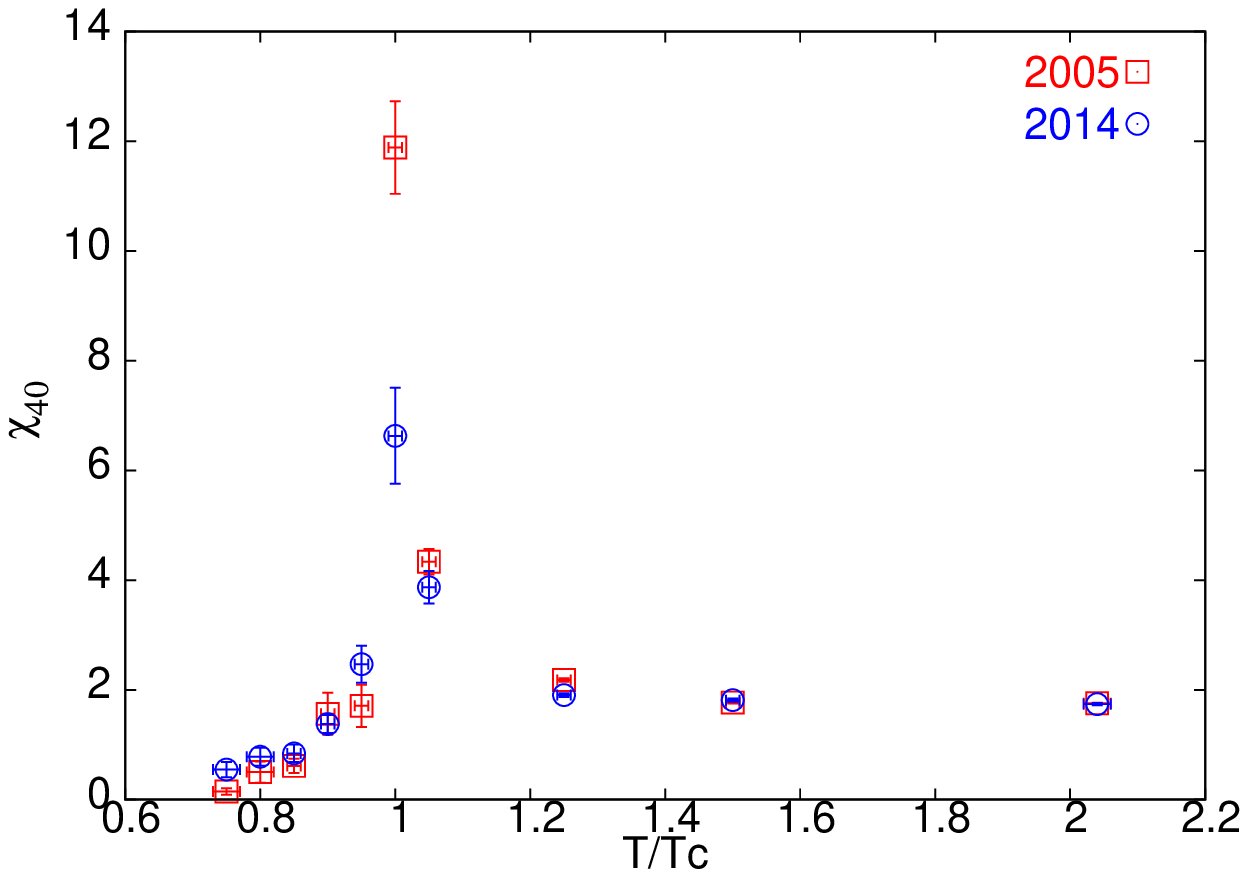}
\caption{Comparison of the new set B and old \cite{nt4} extractions of
 the QNS of order 2 (first panel) and 4 (second panel). The largest
 effects are in the vicinity of $T_c$ in both cases. By decreasing the
 number of source vectors in the analysis, we checked that the lower
 peak value of $\chi_{40}$ in the current runs is due to increased $N_v$.}
\eef{compqns}

\section{Results}\label{sec:results}
\bef
\begin{center}
\includegraphics[scale=0.7]{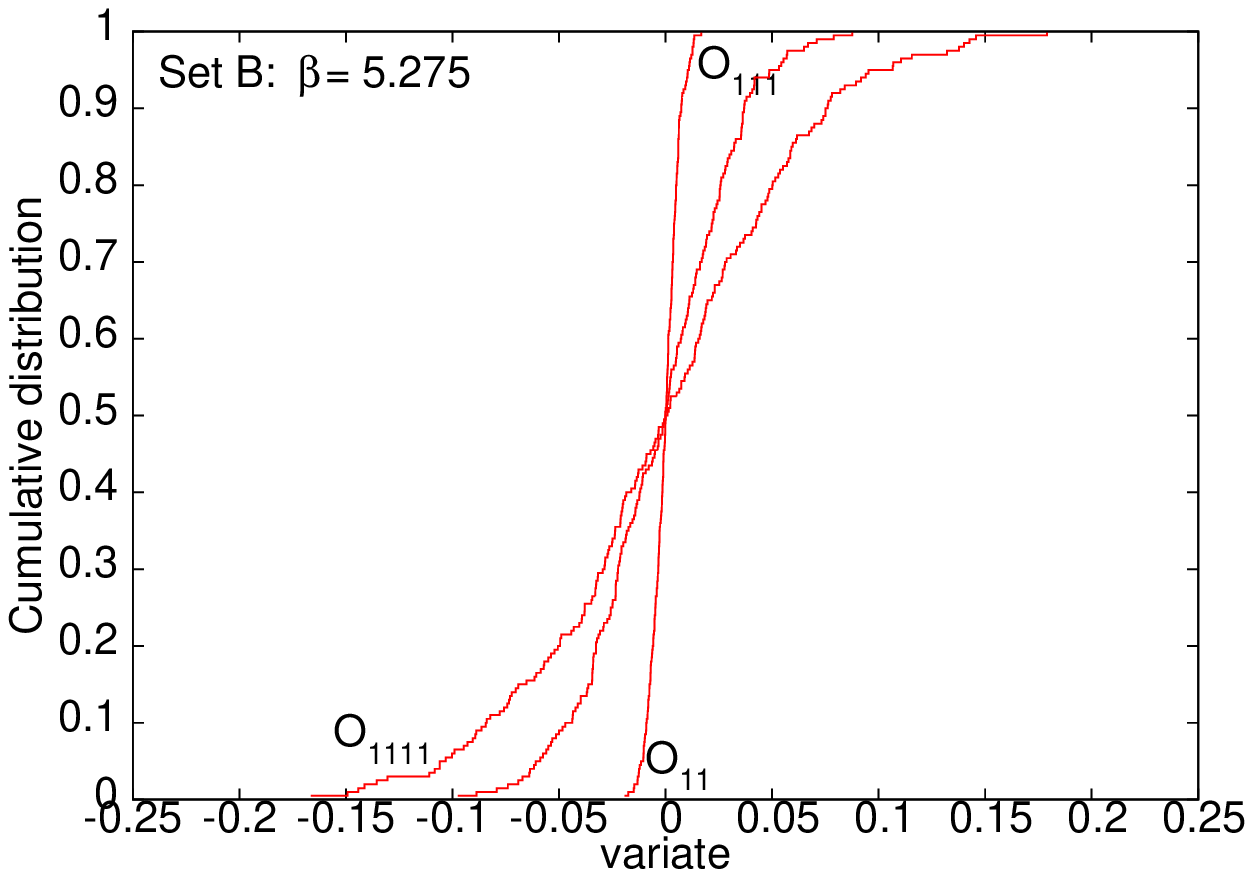}
\includegraphics[scale=0.7]{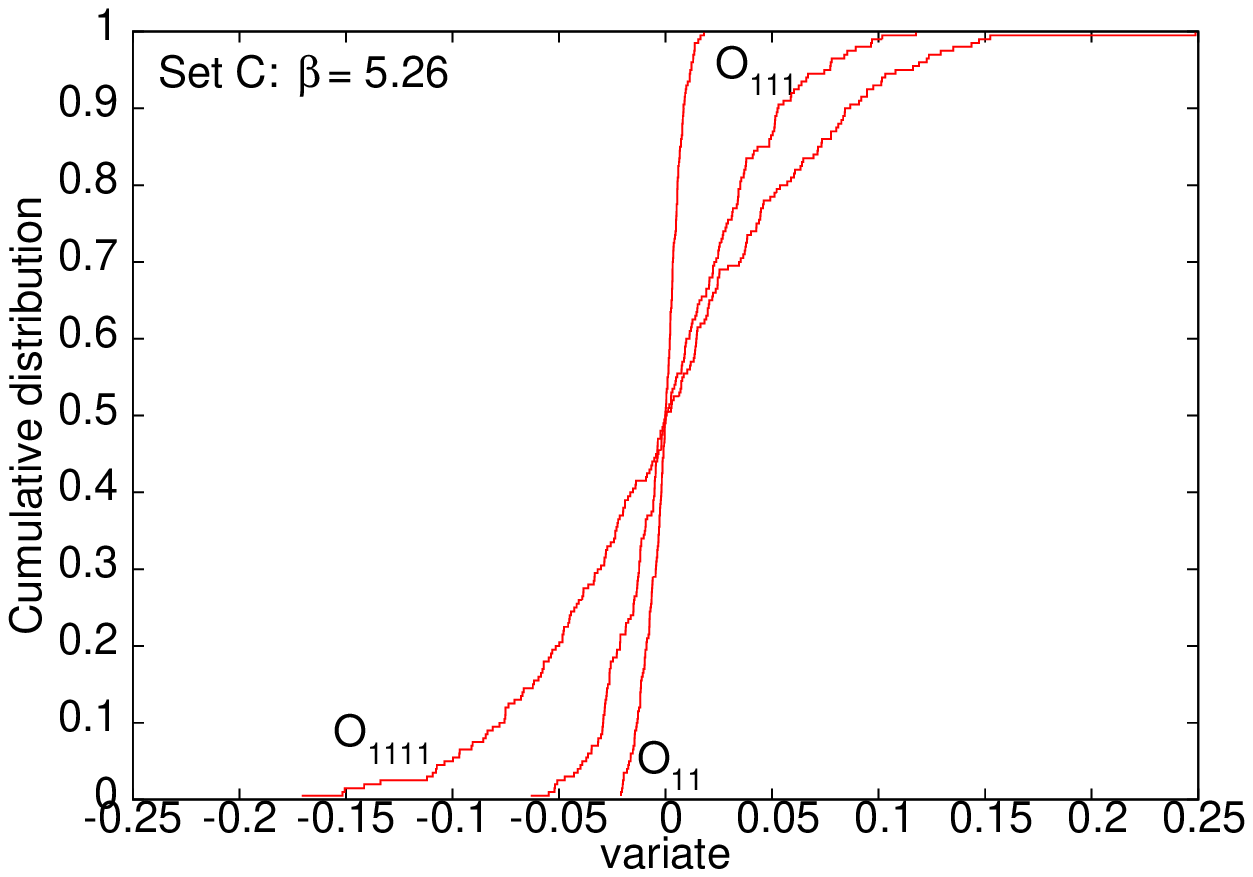}
\end{center}
\caption{Cumulative distributions of ${\cal O}_{11}$, ${\cal O}_{111}$ and
${\cal O}_{1111}$, evaluated by bootstrap and shifted such that the median
 is at zero. Note that, for both values of $m_\pi/m_\rho$, as the number
 of traces in the product increases, the tails of the distribution get fatter.}
\eef{fattail}

\bef
\begin{center}
\includegraphics[scale=0.65]{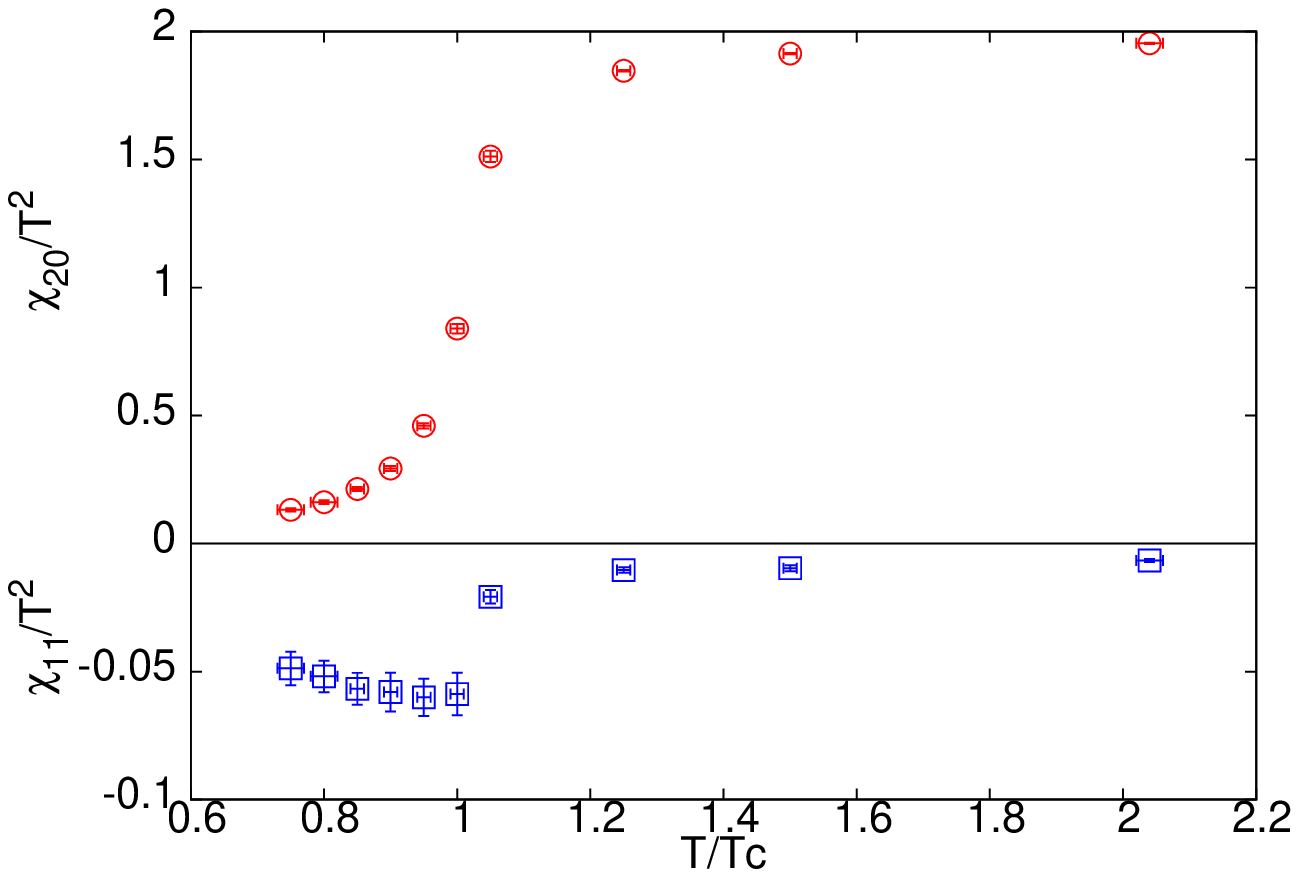}
\includegraphics[scale=0.65]{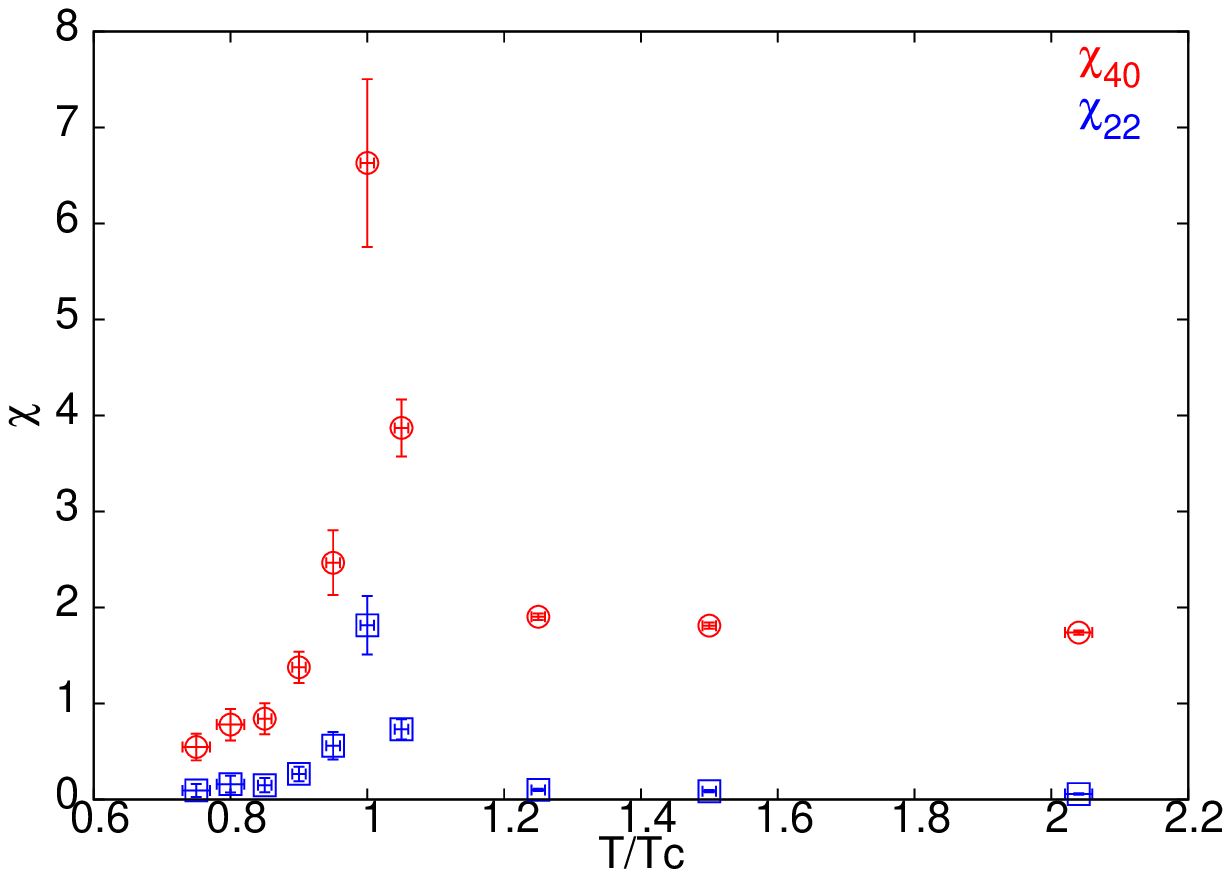}
\includegraphics[scale=0.65]{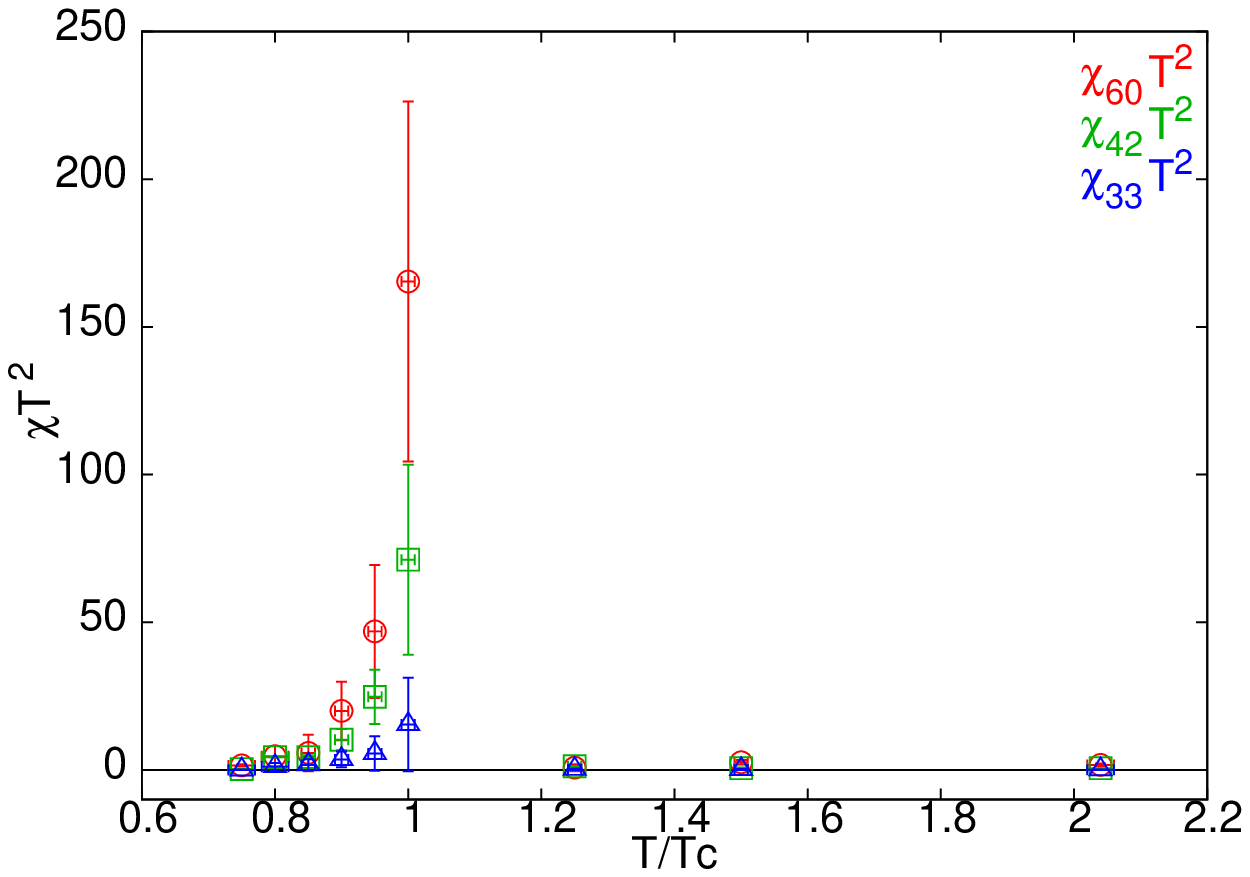}
\includegraphics[scale=0.65]{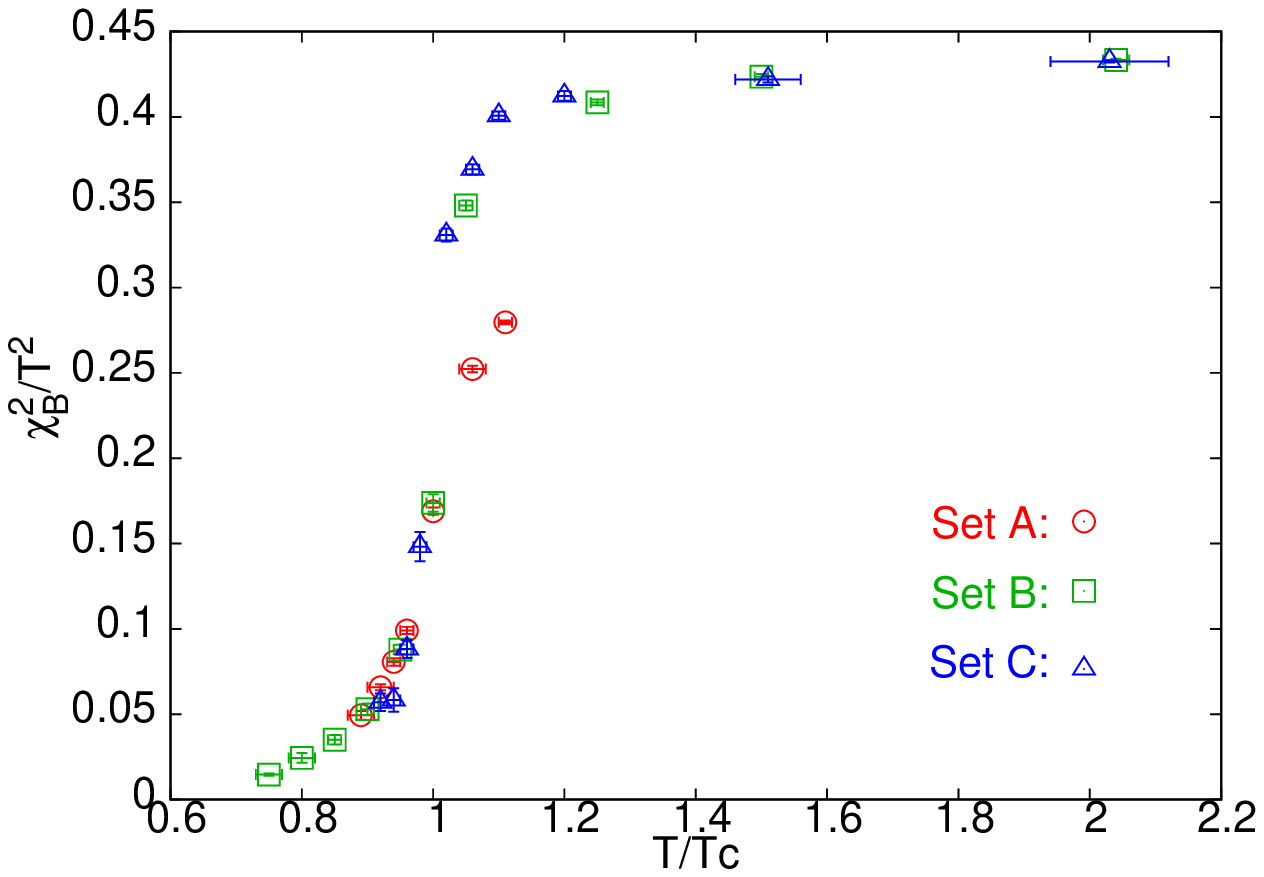}
\end{center}
\caption{QNS in set B. The first panel shows $\chi_{20}/T^2$ (circles)
 and $\chi_{11}/T^2$ (boxes) as functions of $T/T_c$ on $N_t=4$ lattices.
 Notice the difference in scales for the two QNS.
 The second panel shows $\chi_{40}$ (circles) and $\chi_{22}$ (boxes) as
 a function of $T/T_c$ for $N_t=4$ lattices.  The third panel shows
 $\chi_{60}T^2$ (circles), $\chi_{42}T^2$ (boxes) and $\chi_{33}T^2$
 (triangles) as functions of $T/T_c$. The fourth panel shows $\chi_B^2/T^2$
 for the three sets.}
\eef{ord4}

\bef
\begin{center}
\includegraphics[scale=0.7]{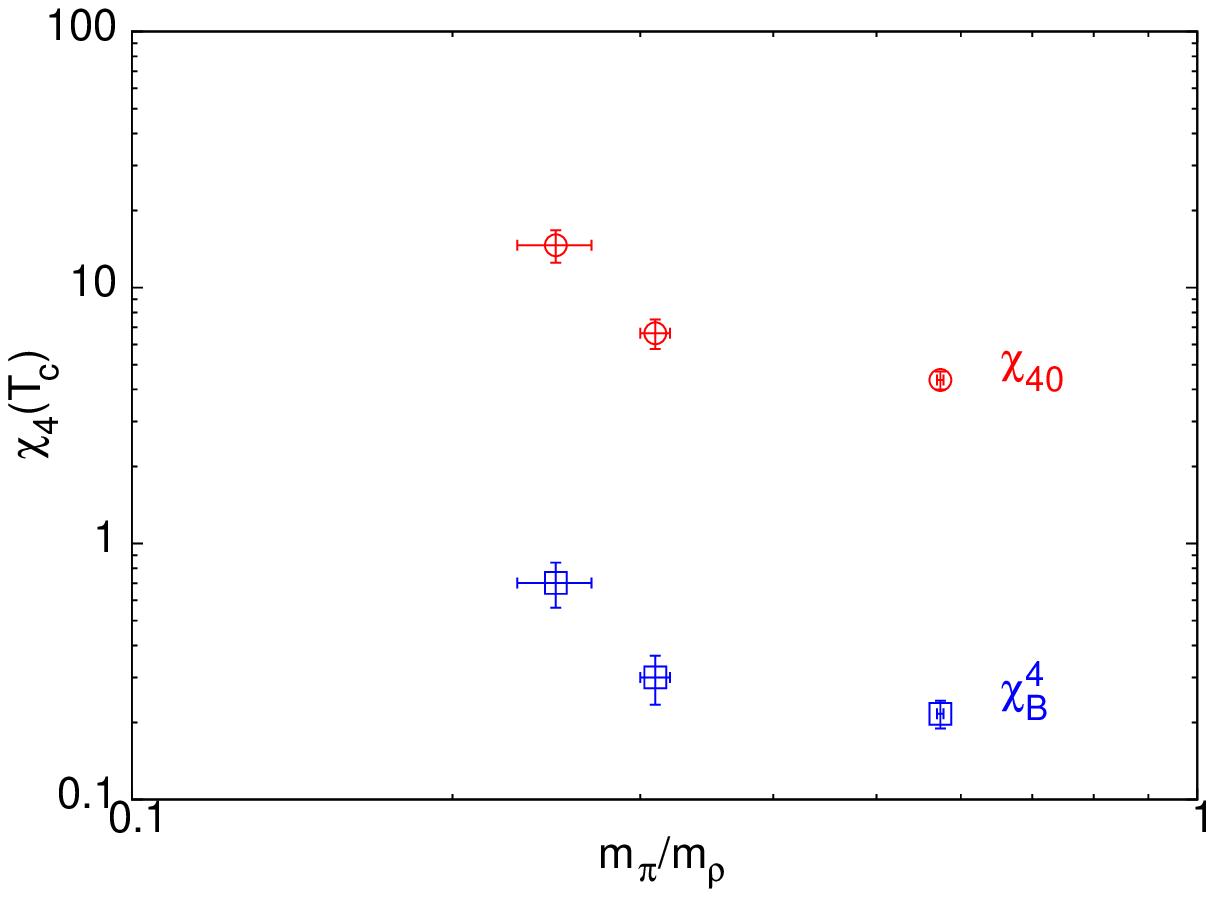}
\includegraphics[scale=0.7]{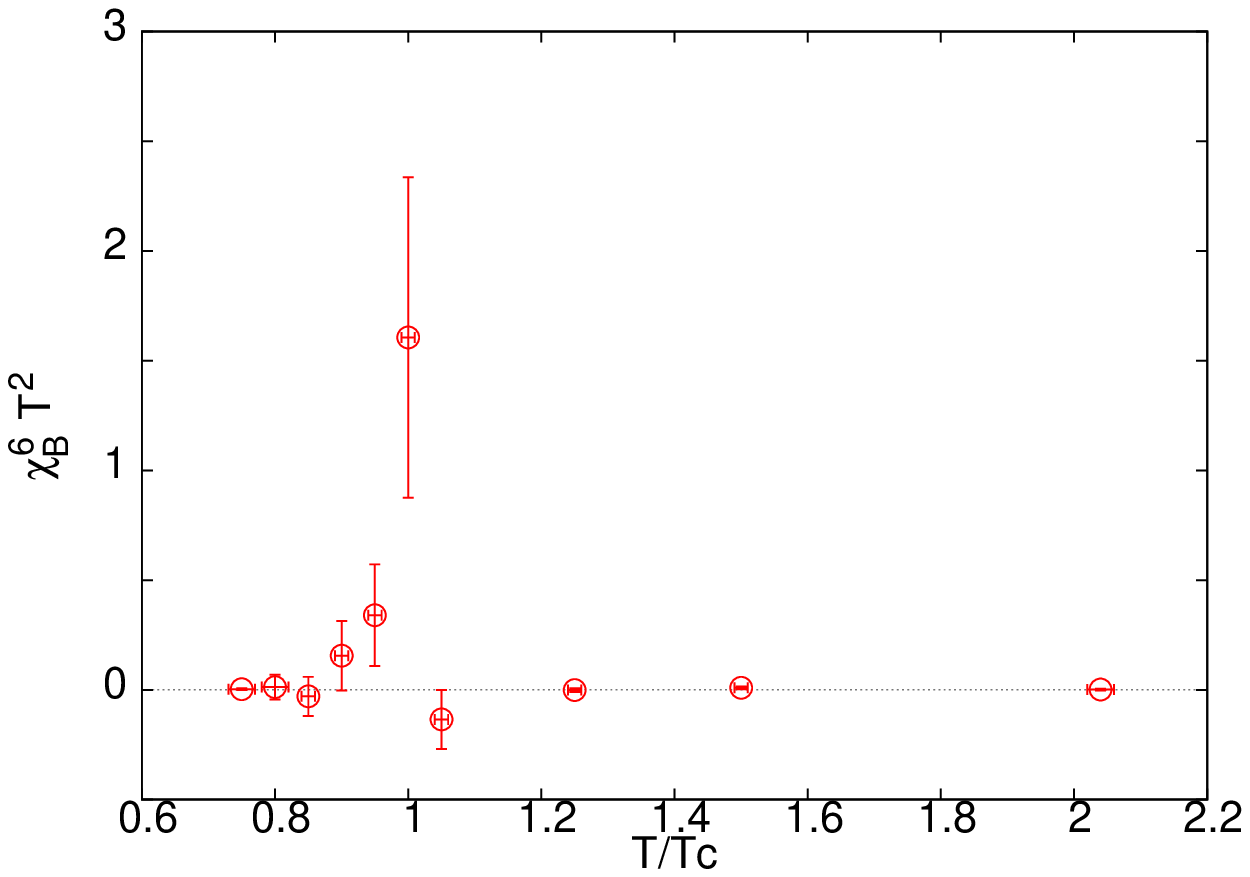}
\end{center}
\caption{The first panel shows the dependence of the peaks of two fourth
 order QNS as functions of $m_\pi/m_\rho$; a power law does not seem to
 be a good descriptions. The second panel shows the dependence of
 $\chi_B^6 T^2$ on $T/T_c$ for set B. A peak is seen at $T_c$, where
 $\chi_B^4$ also peaks.}
\eef{massdep}

\bef
\begin{center}
\includegraphics[scale=0.7]{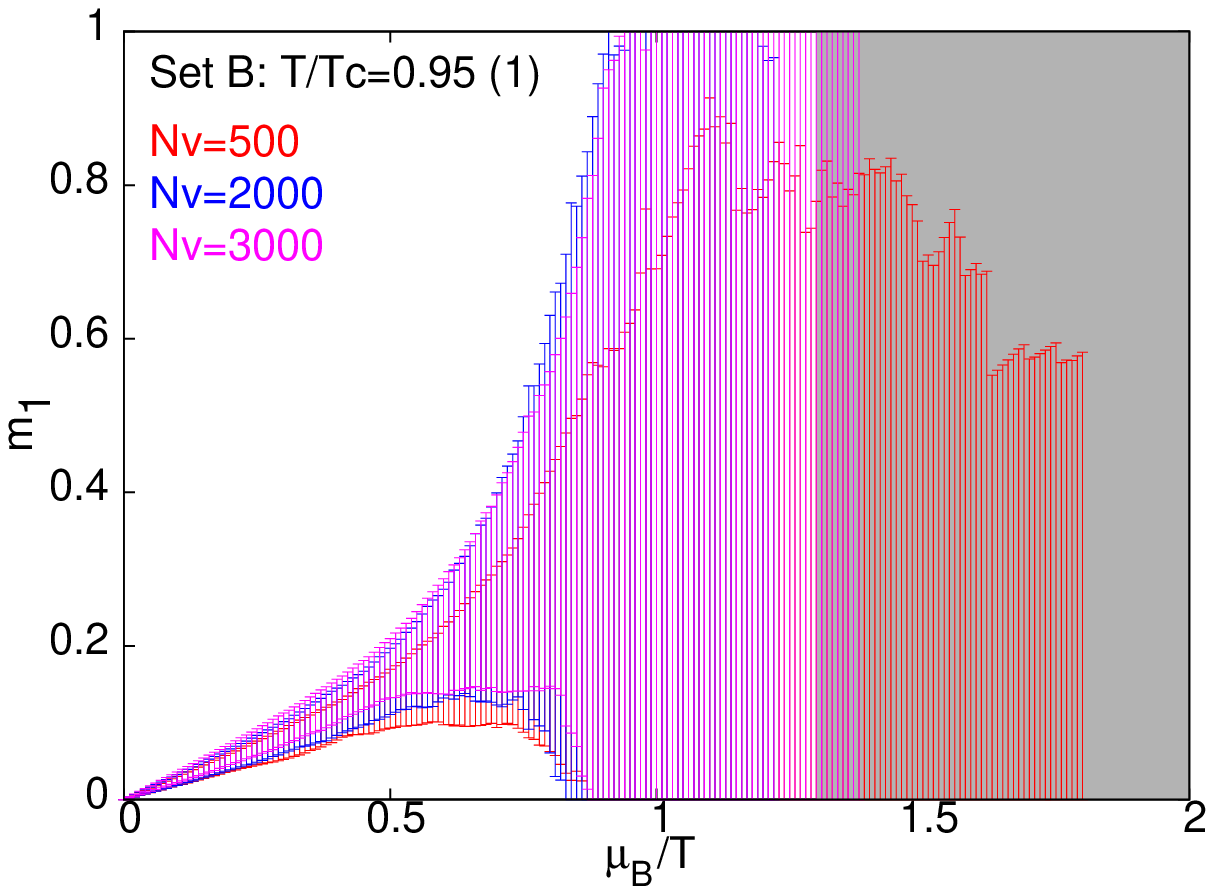}
\includegraphics[scale=0.7]{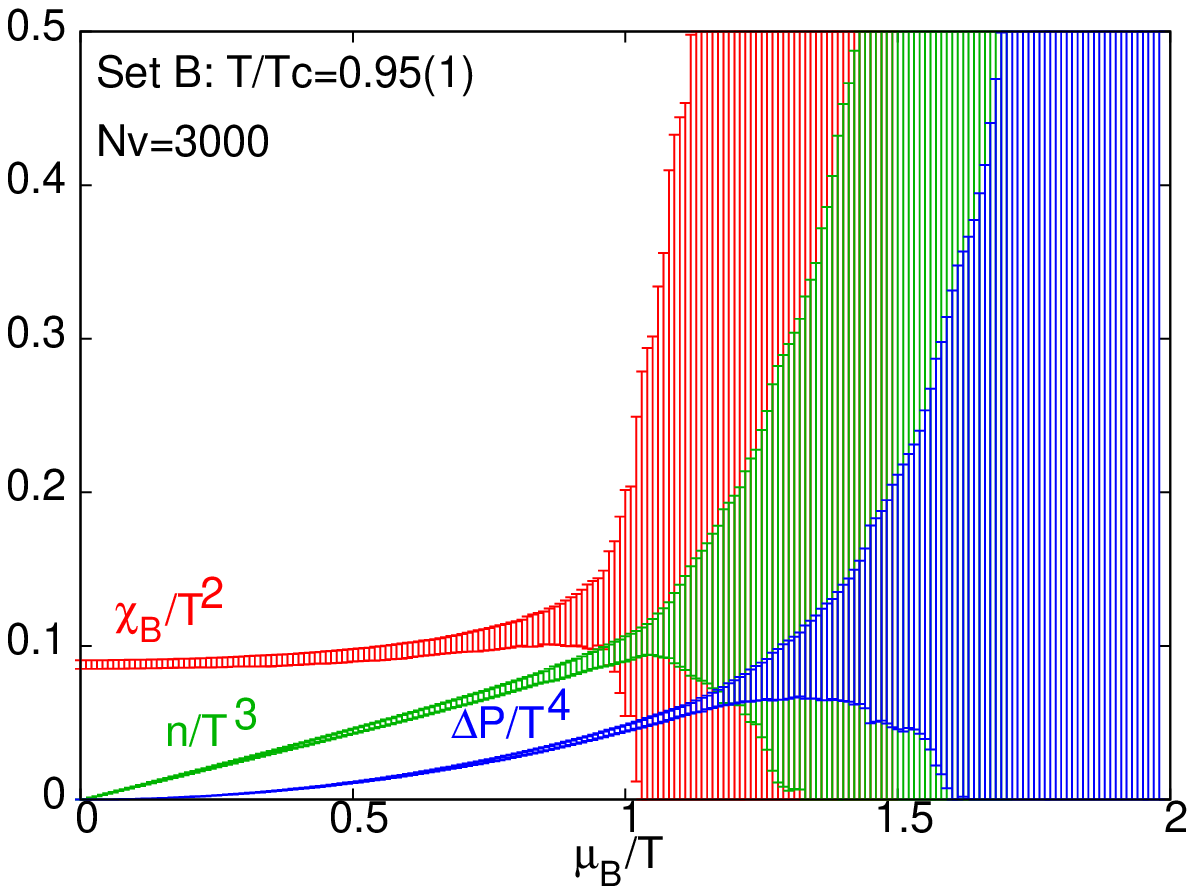}
\end{center}
\caption{The DLOG Pad\'e approximant, $m_1$, and its errors are shown in
 the first panel; the dark band shows the radius of convergence obtained
 by analysis of the series coefficients. By successive
 integration one constructs $\chi_B$, $n$ and $\Delta P$, as shown in the
 second panel. The errors at different values of $\mub/T$ are strongly
 correlated, and the reason for the decrease in errors with integration
 order is discussed in the text.}
\eef{m1}

\bef
\begin{center}
\includegraphics[scale=0.7]{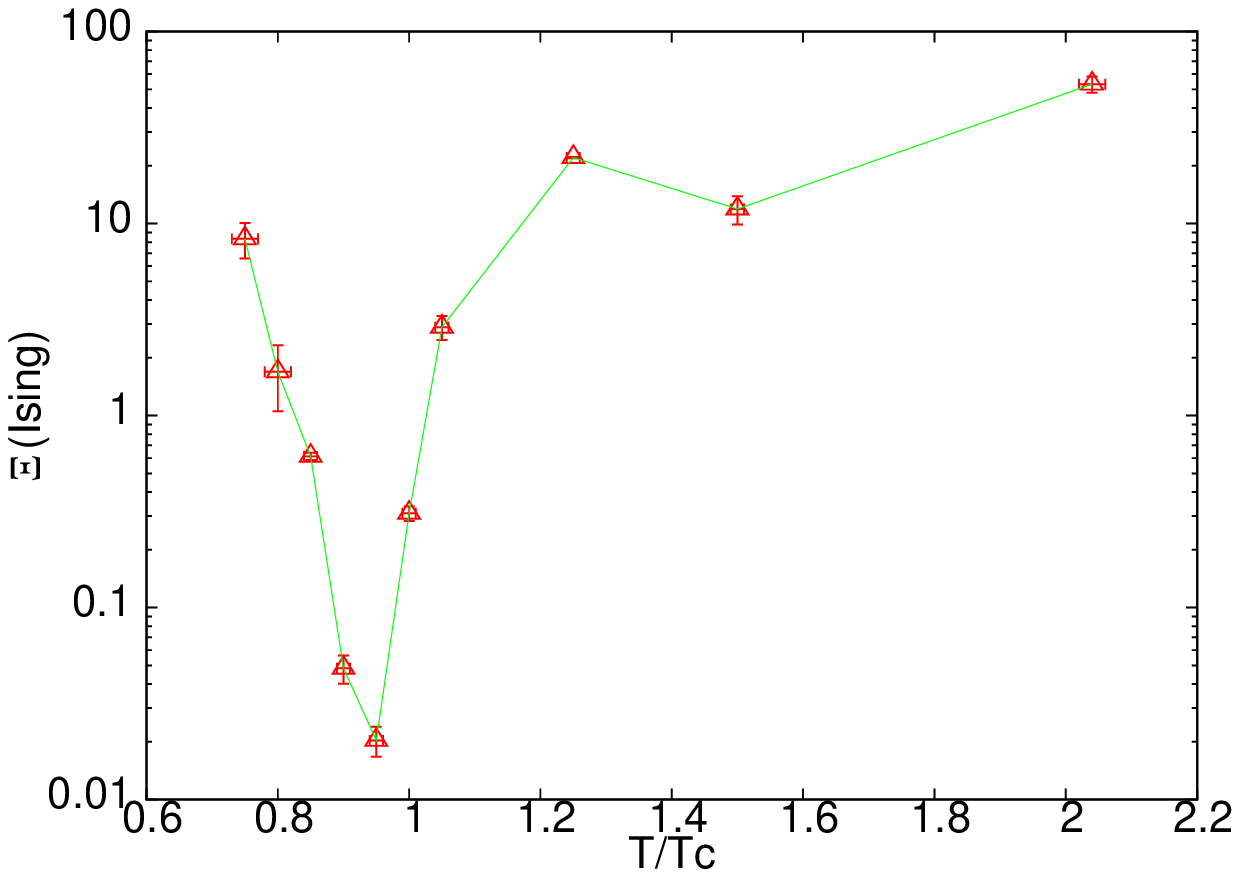}
\includegraphics[scale=0.7]{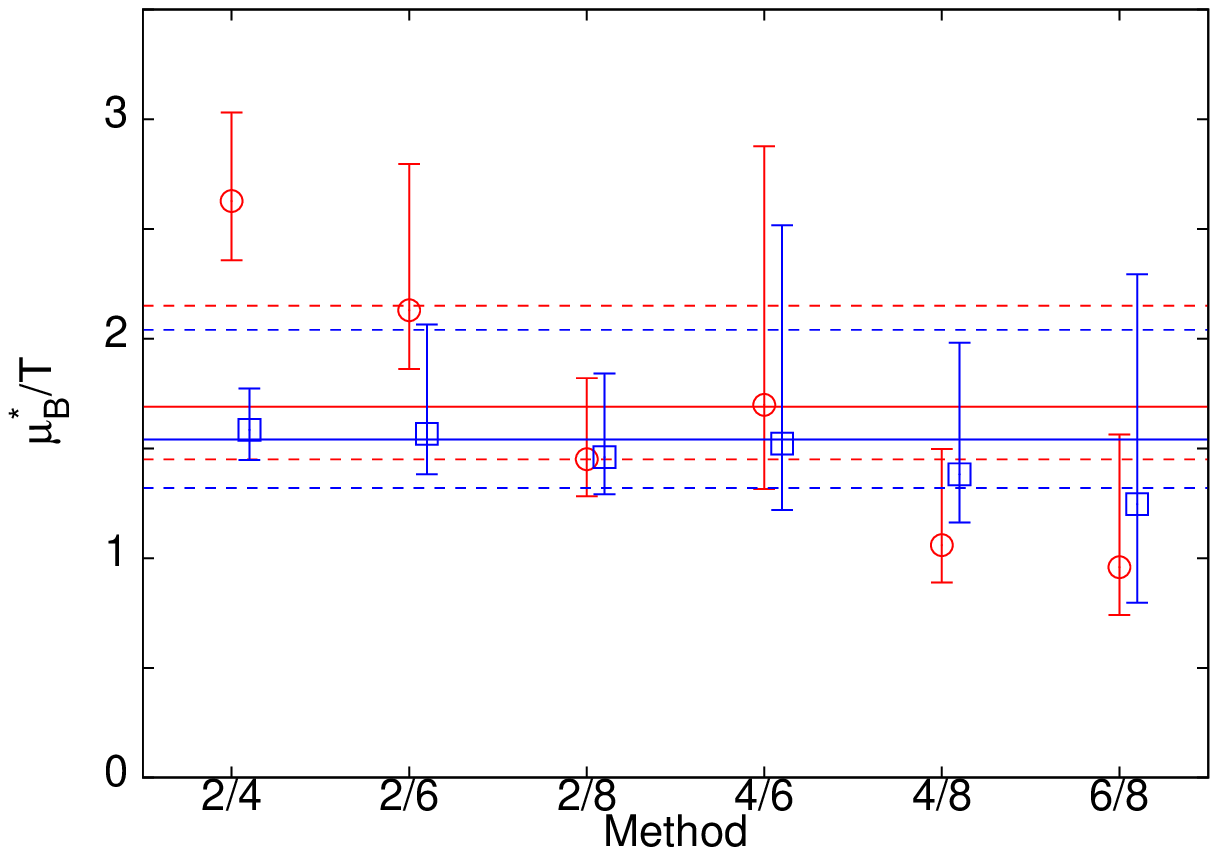}
\end{center}
\caption{The first figure gives a measure of the goodness of the DLOG Pad\'e
 resummation assuming the critical behaviour to be in the Ising universality
 class in set B. We estimate $T^E/T_c$ to be at the minimum of this curve.
 The second figure shows the radius of convergence obtained from the ratios
 of various terms in the series expansion of $\chi_{20}$ (boxes) and the
 successive $\chi_B^n$ (circles), obtained in set B. The best common value
 and its errors are marked by the bands.}
\eef{radconv}

\bef
\includegraphics[scale=0.7]{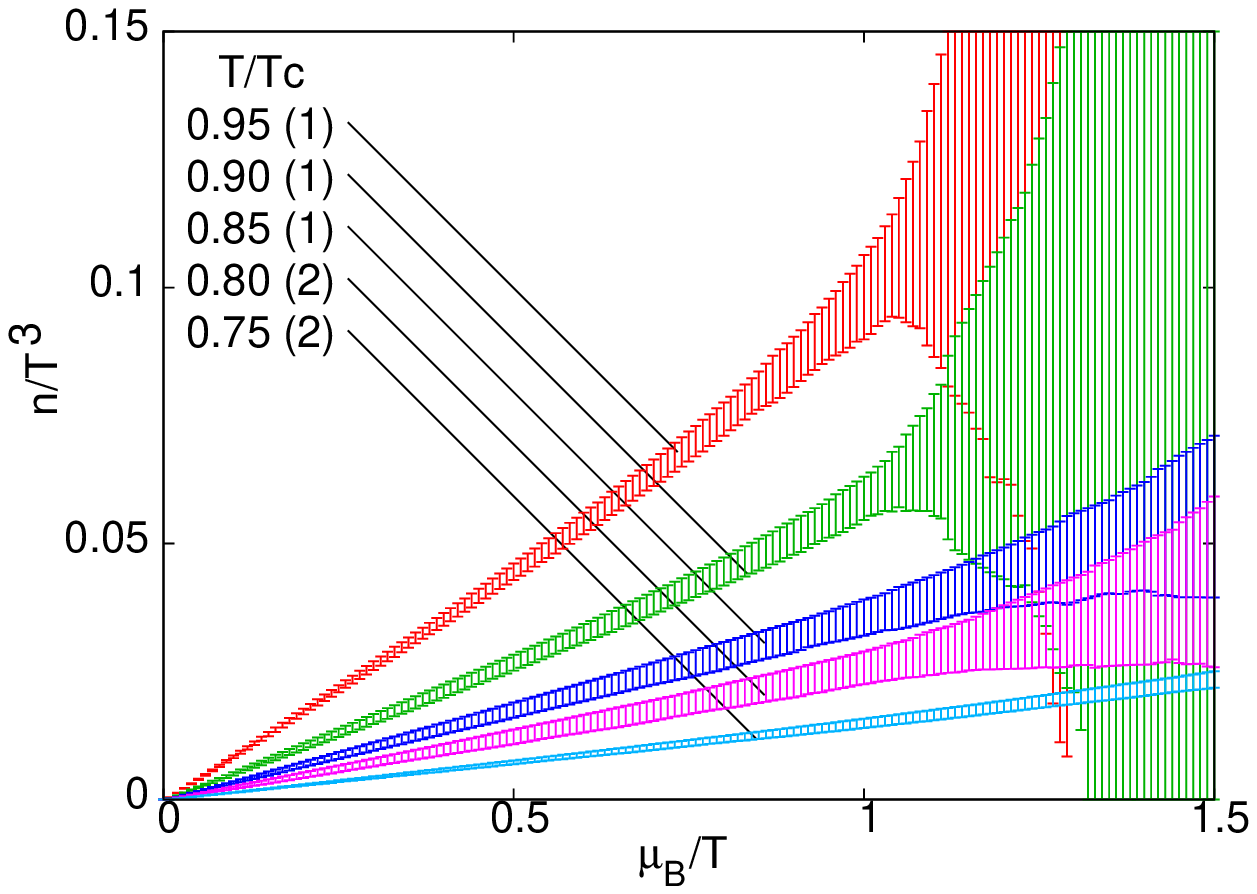}
\includegraphics[scale=0.7]{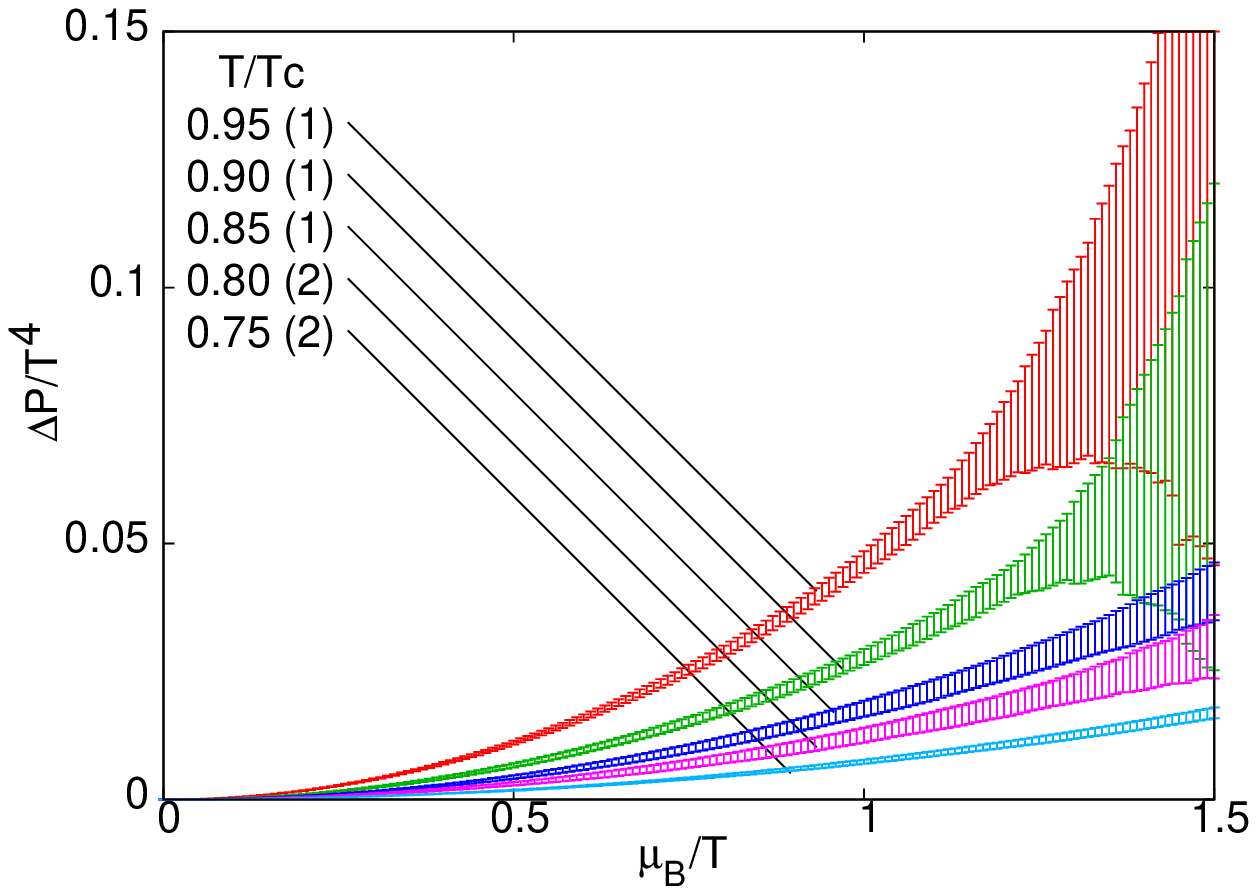}
\caption{The quark number density and the baryon contribution to the
 pressure as a function of $\mub/T$ for various temperatures below
 and up to $T_E$ in set B. The view as a function of $T/T_c$ for several
 different $\mub$ is shown in \fgn{tnpress}.}
\eef{npress}

\bef
\begin{center}
\includegraphics[scale=0.7]{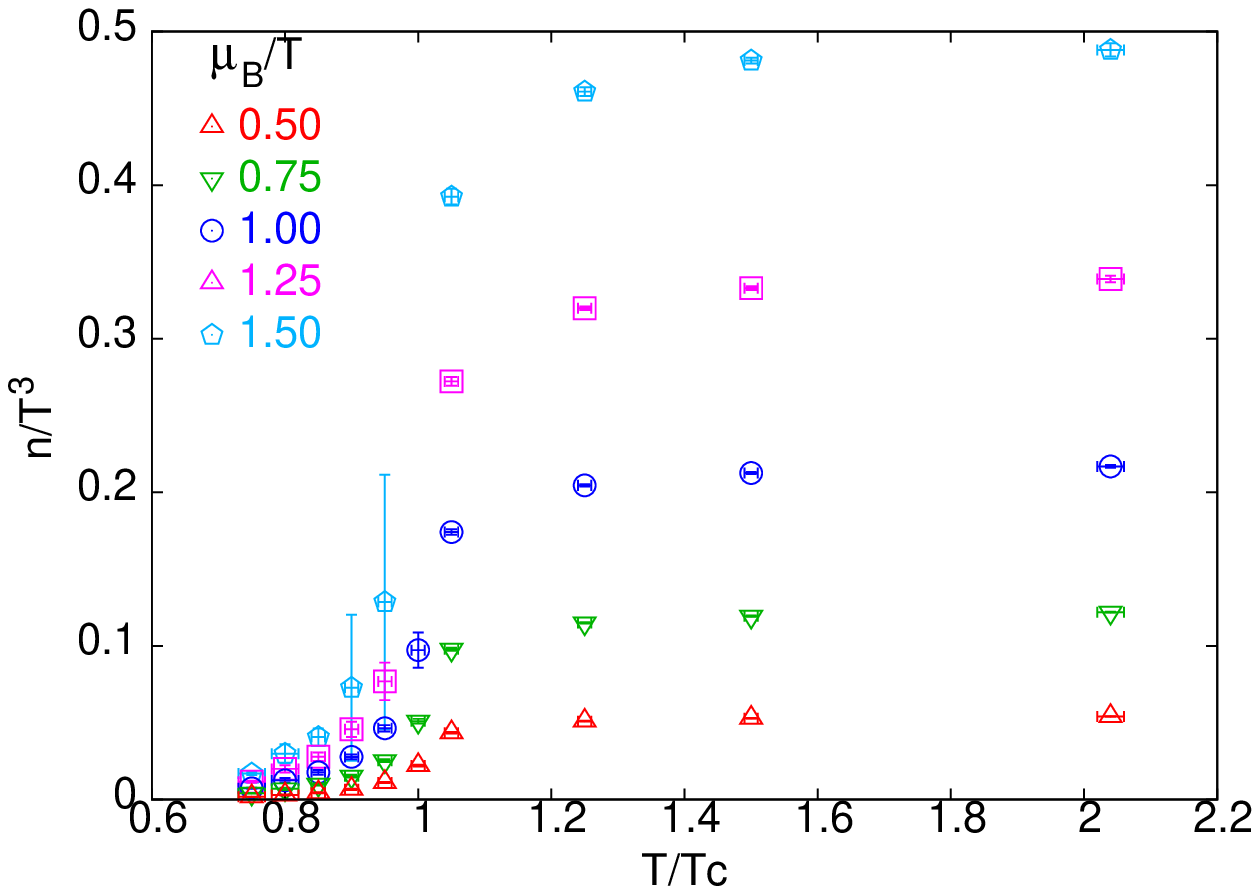}
\includegraphics[scale=0.7]{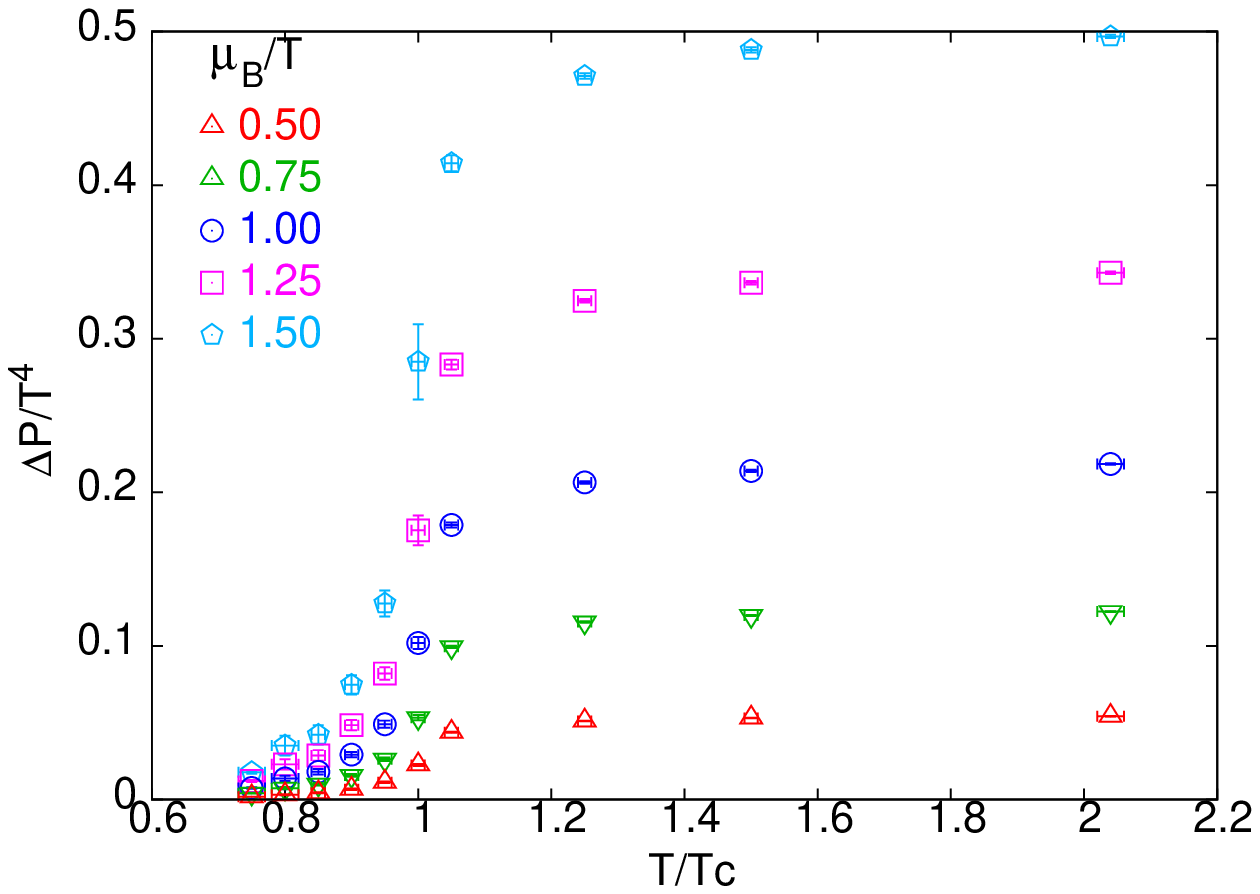}
\end{center}
\caption{The equation of state for set B shown as a function of $T/T_c$
 at various different values of $\mub$. The view as a function
 of $\mub$ at several different $T/T_c$ is given in \fgn{npress}.}
\eef{tnpress}

\bef
\includegraphics{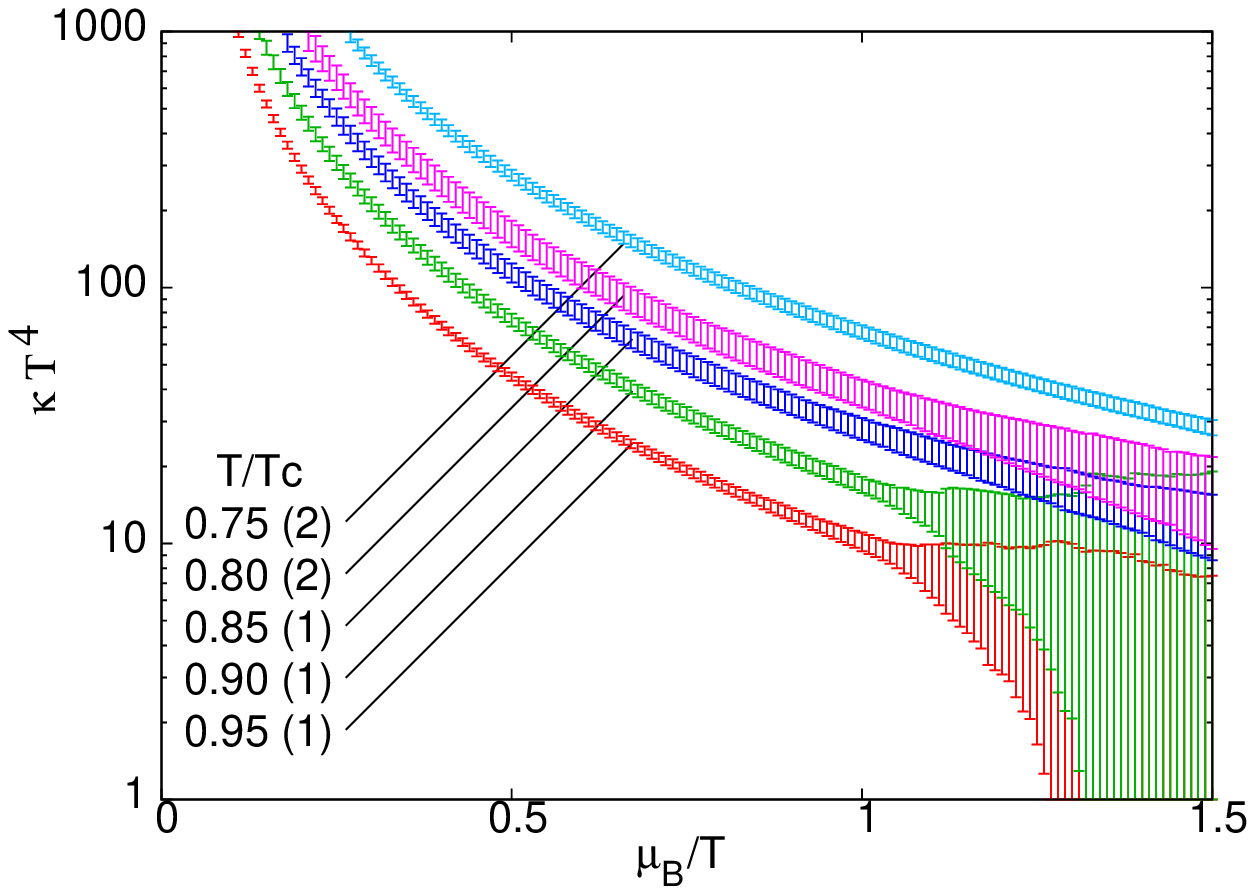}
\caption{The isothermal bulk compressibility in QCD as a function of
 the baryon chemical potential in set B.}
\eef{kappa}

\bef
\includegraphics[scale=0.7]{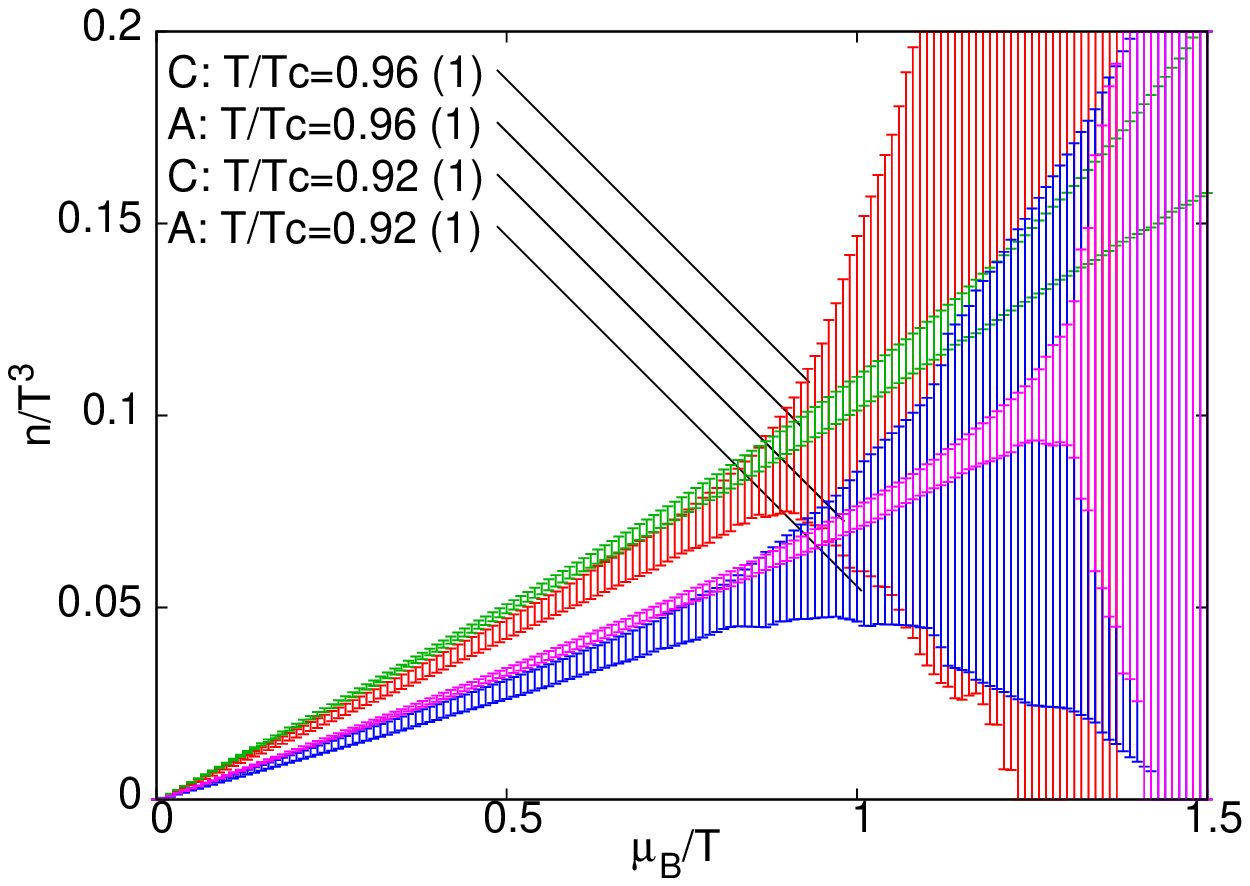}
\includegraphics[scale=0.7]{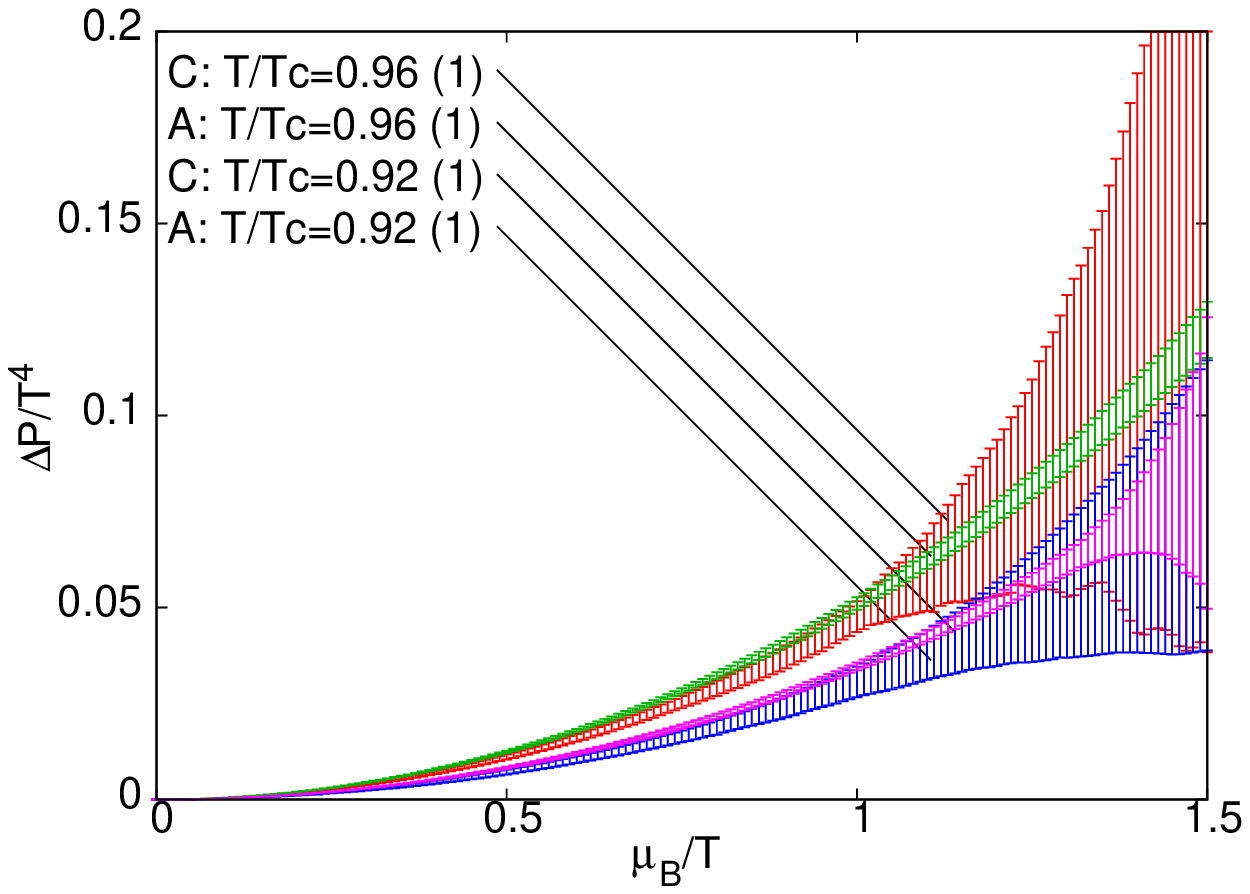}
\caption{The quark number density and the baryon contribution to the
 pressure at different temperatures and two different pion masses at each
 temperature. By comparing the two sets, one sees that the effect of the
 pion mass on the equation of state vanishes within errors over this range
 of pion masses.}
\eef{mpress}

\subsection{Errors in the QNS}

Each QNS is the sum of the expectation values of a set of operators with
different topologies as in \eqn{topol},
\beq
   \chi_i = \sum_{\alpha=1}^M n_{i\alpha}\langle{\cal O}_{\alpha}\rangle
          = \sum_{\alpha=1}^M \langle Q_{\alpha}\rangle = u\cdot Q.
\eeq{qnsops}
We have introduced the notation $Q_\alpha=n_{i\alpha}{\cal O}_\alpha$,
as well as the $M$-dimensional vector $Q$ whose $\alpha$-th component
is $\langle Q_\alpha\rangle$.  The $M$-dimensional vector $u$ has all
components equal to unity.

Some insight into which terms dominate the errors comes from a
pseudo-Gaussian analysis.  At each temperature, the largest components
of $Q$ dominate the expectation of a QNS.  To find which terms
dominate the error, we make a bootstrap estimate of the covariance
matrix, $\Sigma_{\alpha\beta}$ of the $Q_\alpha$; its eigenvalues are
$\sigma^2_\gamma$ (ordered such that $\gamma=1$ is the largest) and
the normalized eigenvectors are $v_\gamma$. If $Q\cdot v_\gamma=|Q|
\cos\theta_\gamma$, then clearly
\beq
   \chi_i = |Q| \sum_{\gamma=1}^M \cos\theta_\gamma u\cdot v_\gamma, 
       \qquad{\rm and}\qquad
   {\rm Var\ \/}\chi_i = |Q|^2 \sum_{\gamma=1}^M \cos^2\theta_\gamma
       (u\cdot v_\gamma)^2 \sigma^2_\gamma,
\eeq{normal}
and the contribution of each term in the second sum defines which is the
largest contributor to the error. The dominant terms in the two sums
may not be the same. The analysis in terms of the covariance matrix
would be exact if the errors were Gaussian. When the distributions of
the variables are non-Gaussian, as here, the analysis still determines
the most fluctuating directions in the neighbourhood of the mean.

Above $T_c$ the contributions to the QNS are similar to that predicted
by simple power counting \cite{bir}. The contributions to the errors are
also dominated by the same operators. The situation below $T_c$ is more
complicated.  Second order QNS are dominated both in magnitude and error
by the contribution from $\langle{\cal O}_2\rangle$, if it contributes.
Fourth order QNS are dominated by the 2-loop operator $\langle{\cal
O}_{22}\rangle$ and the 3-loop operator $\langle{\cal O}_{112}\rangle$,
the precise mix depending on the coefficients $n_{i\alpha}$. The errors
generally come from just one eigenvector: that corresponding to the
largest eigenvalue of the covariance matrix. In this, the 3-loop operator
generally dominates, although the 4-loop $\langle{\cal O}_{1111}\rangle$
has a significant contribution.  Sixth order QNS have significant
contributions from $\langle{\cal O}_{222}\rangle$ and $\langle{\cal
O}_{1122}\rangle$, the hierarchy being decided by the $n_{i\alpha}$.
The errors are due essentially only to the leading eigenvector,
which has major contributions from the $\langle{\cal O}_{1122}\rangle$,
$\langle{\cal O}_{11112}\rangle$ and $\langle{\cal O}_{111111}\rangle$.
Although a significant fraction of the error comes from operators whose
contribution to the expectation is negligible, it does not seem possible
to improve the errors except by throwing out ``small'' terms. Since this
may introduce systematic errors, we do not explore this drastic solution.

The origin of large errors lies in the fact that the multi-loop
operators are the products of individual loops. The measurement of
each loop yields a random number with a distribution which depends on
the bare parameters of the simulation. The distributions of products
of random variates typically are non-Gaussian and have fat tails; we
show some examples in \fgn{fattail}. Clearly more loops lead to fatter
distributions. Unfortunately, distributions of products of random
numbers have not received much attention in the literature. Examples
in which the random variates are positive have been examined recently
\cite{kaplan}. The case at hand is more complicated since the variates
are of indefinite sign.

\subsection{The QNS and chiral critical behaviour}

The second order QNS are shown in \fgn{ord4}. Smooth variations are seen
as functions of $T/T_c$ for both the QNS. The errors in the measurement
of $\chi_{11}/T^2$ are much reduced compared to previous measurements. As
a result, we can clearly see a smooth temperature variation.  Two of the
fourth order QNS are also shown in \fgn{ord4}. Each has a sharp
peak at $T_c$. We can trace this to a peak in $\langle\op{22}\rangle$
at $T_c$. We also find a sharp peak at $T_c$ in the sixth order QNS,
which is exhibited in \fgn{ord4}. This peak is not a statistical artifact
since it is develops into a stable value as $N_v$ is increased to 2000. 

The variation of the second order QNS with mass is displayed in the fourth
panel of \fgn{ord4}. Below $T_c$ the dependence of $\chi_B^2/T^2$ on
$m_\pi$ seems to be statistically insignificant. At $T_c$ and immediately
above it, there is some dependence of this QNS on the quark mass. For set
A $m_\pi/T_c>3$, so the mass effect in the high temperature phase seems
to be strong, and is likely to persist until fairly high $T$. For set B,
$m_\pi/T_c$ is less than half of this value, and we see that at $1.5T_c$
sets B and C give similar results. The fourth order QNS have the form
shown in \fgn{ord4}, with a peak at $T_c$. We find little mass dependence
below $T_c$, and some at $T_c$ and above, although sets B and C coincide
for $T\ge1.5T_c$.

It was conjectured earlier that the chiral critical behaviour for
$m_\pi=0$ and $\mub=0$ yields a scaling form, $f_s(t)$, for the singular
part of the free energy as a function of the variable
\beq
   t = a \left(\frac T{T_c}-1\right) + b \left(\frac{\mub}{T_c}\right)^2,
\eeq{fkrscaling}
where $a$ and $b$ are scale parameters \cite{friman}. Successive
derivatives of $f_s$ with respect to $T$ are then proportional to successive
double derivatives with respect to $\mub$, \ie
\beq
   \chi_B^{2n}(T)= \left.\frac{\partial^{2n}f_s}{\partial\mub^{2n}}
       \right|_{\mub=0}
   \propto\left.\frac{\partial^nf_s}{\partial T^n}\right|_{\mub=0}.
\eeq{derivs}
A sharp rise in $f_s$ across the chiral phase transition would then give
a peak at $T_c$ in $\chi_B^4$ and an oscillatory $\chi_B^6$ with a zero
at $T_c$. Moving away from the chiral limit would still give similar, but
rounded and shifted, features. Such observations would then connect the
finite temperature chiral transition with the QCD critical point. Such
a scaling formula would relate the fourth order QNS to the specific
heat, and so predict a power-law rise of $\chi_B^4$ as $m_\pi/m_\rho$
decreases. The power can be read off from the formalism presented in
\cite{friman}.

This important conjecture has not been subjected to a direct test before.
A first test is possible now since we have taken data with several
values of $m_\pi/m_\rho$.  In the first panel of \fgn{massdep} we plot
our measurements of the peak of $\chi_B^4$ and $\chi_{40}$ against
$m_\pi/m_\rho$. A power law is not seen. One possible reason is that
the peak of $\chi_B^4$ is not related to $f_s$. For example,
at the masses we studied, it could be that the free energy is not
dominated by its singular chiral part, $f_s$, and a power law may emerge
at smaller quark masses. In this case a future computation at realistic
pion masses, $m_\pi/m_\rho=0.18$, and smaller, would be needed to test
\eqn{fkrscaling}.  Alternatively there could be finite-volume shifts
in the peaks of $\chi_B^4$ which spoil the apparent scaling. A future
finite size scaling study would then be needed to test this alternative.

In the second panel of \fgn{massdep} we show $\chi_B^6$ in set B,
since it has the best statistics. We see a peak at $T_c$ instead of the
zero which would be predicted by \eqn{derivs}. By changing $N_v$ in our
analysis we have checked that the peak is stable and not a statistical
fluke. This observation also indicates that, in the mass range which
we have explored, the assumption that the free energy is dominated by
the singular term in the chiral limit, with the scaling variable of
\eqn{fkrscaling}, is not supported. Clearly, these first tests should
be supplemented by more detailed studies in future.

\subsection{The logarithmic derivative and its errors}

When the series expansion of a function has a finite radius of
convergence, then summing the truncated series has uncontrolled systematic
errors. One needs a method of resumming the infinite series which remains
reliable even as the statistical errors decrease. Near the critical point
$\chi_B/T^2$ is expected to diverge as $|\mub^2-(\mube)^2|^{-\psi}$.
It turns out to be simpler to examine the logarithmic derivative of
$\chi_B$,
\beq
   m_1(T,\mub) = \frac{\partial\log\chi_B}{\partial\mub}
     \simeq \frac{2\psi\mub}{\mub^2-(\mube)^2}.
\eeq{m1}
The last expression comes by substituting the critical form of $\chi_B$
into the definition. The quantity $m_1$ was initially introduced in
this form to connect with experimental measurements of event-to-event
fluctuations in the baryon number in heavy-ion collisions \cite{cpod,plb}.
The series resummation consists of converting the series for $\chi_B$
into one for $m_1$, and then matching the simple-pole ansatz to it.
This is known as a DLOG Pad\'e approximant in the literature \cite{dlog}.
The scaling assumption, \ie, a single power-law divergence of $\chi_B$ may
turn out to be oversimplified; corrections to scaling can also be
incorporated into the analysis if future improvements in statistics make
it necessary \cite{dlog}. 

Statistical errors in the evaluation of $m_1$ are described in
\apx{pade}, where it is shown that a bootstrap procedure can give good
control of errors except in the neighbourhood of the critical point.
It is shown that with increasing statistics, $N$, one can only control the
errors within a distance ${\cal O}(1/\sqrt N)$ of the critical point. The
need of large statistics to control errors near a critical point is known
as critical slowing down \cite{csd}. This observation reveals that the
use of series expansions also suffers from this generic disease near a
critical point. 

In the first panel of \fgn{m1} we show the resummation of $m_1$ for
different number of source vectors, $N_v$. The effect of increasing
statistics is most clearly visible in the decrease of errors in $m_1$
for $\mub/T<0.75$. For larger values of $\mub/T$ it seems that substantially
larger statistics will be needed to gain control of errors. An estimate can
be made as follows. We find that $m_1$ has 50\% errors at $\mub/T=0.66$
with $N_v=2000$ and at $\mub/T=0.73$ with $N_v=3000$. Then, using the scaling
formul{\ae} of \apx{pade}, this would mean that we would need $N_v>10000$ to
reduce the errors to 50\% at $\mub/T=1$, if we assume that $\mube/T^E=1.5$.

The second panel of \fgn{m1} shows the resummation of the QNS and the
equation of state using the DLOG. The QNS is obtained by integrating
and then exponentiating $m_1$. Then $n$ and $\Delta P$ are obtained
by further integrations. The initial conditions needed are provided by
the measurement of $\chi_B$ at $\mub=0$ and by the conditions that $n$
and $\Delta P$ vanish at $\mub=0$. We use the trapezoidal rule for these
integrations, and tune the integration step size.

If the errors in $m_1$ were simple uncorrelated point-wise errors, then
the integration error would be just the sum in quadrature. This would
result in large errors in $\chi_B$, $n$ and $\Delta P$ close to the
critical point. However, the errors in $m_1$ come from the parameters.
In order to understand errors in the integrals, recall that all the
quantities are functions of $|\mub^2-(\mube)^2|$, and are successive
integrals of $m_1$. So the error, $\delta\Delta P$, in $\Delta P$ can
be written as
\beq
   \left(\frac{\delta\Delta P}{\Delta P}\right)^2 \simeq n^2(\delta\mube)^2,
\eeq{errorprop}
if we assume that the error, $\delta\mube$, in $\mube$ gives the dominant
contribution to $\delta\Delta P$. Since $n$ is regular at the critical
point, so is this error.  These arguments indicate that as we take more
integrals of $m_1$, the errors decrease, since the critical singularity
becomes milder. This is clearly visible in \fgn{m1}.

Also visible in \fgn{m1} is the fact that integration smooths out
singularities. The divergence of $m_1$ is clear, the milder divergence
of $\chi_B$ is less so, and the subtle changes in the slopes of $n$
and $\chi_B$ would take significantly more statistics to see clearly.
Indeed, it is through the abrupt increase in errors, related to critical
slowing down, that one can infer most easily the presence of a critical
point.

\subsection{Critical behaviour}

Analysis of the series coefficients in the expansion of $\chi_{20}$
gives the radius of convergence. When the series coefficients are all
clearly positive, then the radius of convergence corresponds to a real
singularity. It turns out that in all three sets of runs, the Taylor
series at $T/T_c\simeq0.95$ has a rather clear signal of a singularity
on the real $\mub$ axis. This is confirmed by the DLOG Pad\'e analysis.

The DLOG Pad\'e analysis of the series expansions of $\chi_{20}$ and
$\chi_B$ agree with each other. For set B we find that the poles lie
off the real axis for $T<0.9T_c$ and $T>T_c$. In these temperature
intervals $\psi$ turns out to be negative. So, there is no critical
point outside the range $0.9T_c\le T\le T_c$. Within this range the
series expansion for the DLOG is best compatible with the single pole
ansatz for $T/T_c=0.95\pm0.01$.

From the residue of the DLOG Pad\'e at the pole one should be able
to extract the critical index. Unfortunately, the errors are much too
large for us to be able to quote a value. However, at our estimate of
the critical point we obtain a value of $\psi$ compatible with that
expected in a 3-d Ising model as well as the mean-field theory (see
\apx{widom}). Also, at this point, if we constrain the critical index to
its Ising value, the location of the pole of the resulting DLOG Pad\'e
approximant is compatible with the estimate from the ratios of Taylor
coefficients.

An useful confirmatory test is the following. Assume that the series
coefficients of the DLOG determined numerically from the QNS are $D_i$.
Then equating this to the Pad\'e approximant gives
\beq
  D_0\mu + D_1\mu^3 + D_2\mu^5 + D_3\mu^7 + \cdots
     \simeq \frac{2\psi\mub}{\mub^2-(\mube)^2},
\eeq{approx}
Since there are only two parameters to be determined for the right hand
side, a series of more than two terms for $m_1$ can test whether the pole
is a good ansatz for the resummation of the series for $m_1$. Doing this
amounts to making a series expansion of the right hand of \eqn{approx}
by synthetic division, yielding a series with coefficients $P_i$. Then one
can check whether $D_i-P_i$ vanishes within errors. A simple statistic for
this test is 
\beq
   \Xi=\frac{E(D_i-P_i)^2}{\sigma(D_i-P_i)^2},
\eeq{xidef}
where $E$ denotes the expectation value, and $\sigma$ the error.
In \fgn{radconv} we show $\Xi$ obtained with the second term of the
series for set B, when the first term is used to obtain $\mube$ assuming
that $\psi=0.79$.  Smaller values of $\Xi$ indicate a good fit.  If the
distribution of the $D_i$ and $P_i$ were Gaussian, then $\Xi$ would be
$\chi^2$-distributed, and it would be natural for a good fit to have
$\Xi\simeq1$. However, both $D_i$ and $P_i$ are fat-tailed and strongly
non-Gaussian, as we have discussed before, so the small values of $\Xi$
at the minimum are not unnatural.

As one can see in \fgn{radconv}, the series expansion in set B indicates
that the critical region lies in the range of temperatures $0.9T_c\le T\le
T_c$, with $T/T_c\simeq0.95$, being the most probable location of the
critical point.  The same range is selected out by the DLOG analysis of
$\chi_B$ as well as $\chi_{20}$, and whether $\psi$ is fixed or allowed
to vary.

\bet
\begin{tabular}{|c|cc|cc|}
\hline Set & $m_\pi/m_\rho$ & $m_N/m_\rho$ & $T^E/T_c$ & $\mube/T^E$ \\ \hline
A & $0.574\pm0.004$ & $1.61\pm0.01$ & $0.94\pm0.02$ & $1.5^{+0.2}_{-0.1}$ \\
B & $0.32\pm0.03$ & $1.57\pm0.08$ & $0.95\pm0.01$ & $1.5^{+0.5}_{-0.2}$ \\
C & $0.25\pm0.02$ & $1.4\pm0.1$ & $0.96\pm0.01$ & $1.4^{+0.4}_{-0.2}$ \\ \hline
\end{tabular}
\caption{The position of the critical end point obtained using methods
 discussed in the text for the three sets of runs we use in this paper.
 The ratios of hadron masses are taken from \cite{oldmilcm,nikhil}.}
\eet{cep}

Given these multiple indicators we estimate the position of the critical
point in the following manner. First we find the temperature at which
the resummation works best, using the $\Xi$ measure introduced above.
We do this with fixed and floating $\psi$.  In each of the three sets
at hand, the best temperature obtained by these tests coincides. This
gives us $T^E/T_c$. At this temperature we examine whether the series
coefficients are all positive; again in all three sets we find that
they are. Then we use the ratios of various coefficients in the series
expansion, as done in earlier computations, to obtain an estimate of
the radius of convergence, $\mube/T^E$. We check that this is consistent
with the position of the pole in the DLOG Pad\'e analysis, as they are
in the three sets we examine.  The results are collected in \tbn{cep}.

There is no change in $T^E/T_c$ obtained for set B when compared
with the earlier results in \cite{nt4}, however there is a shift in
$\mube/T^E$ from the value $1.1\pm0.2$ reported earlier to the value
given in \tbn{cep}.  By repeating the analysis with $N_v=100$ we obtain
results near the older value. This leads us to conclude that the shift
is due to the increased $N_v$ used in the current study. Over the range
of masses we explored, a mild shift in the critical end point towards
$T_c$ and smaller $\mu$ cannot be ruled out by the data. However, simple
linear and quadratic fits show that extrapolations to the physical
point $m_\pi/m_\rho=0.18$ do not change the location of QCD critical
point significantly.

\subsection{The equation of state}

Our results in set B for the bulk thermodynamic quantities, $n$ and
$\Delta P$, are shown in \fgn{npress}. Both these bulk quantities increase
monotonically with $T$ and $\mub$, as expected. Away from our estimated
critical point the errors are small. Even so, the truncated sum is
consistent with the resummed value. As one approaches the critical point,
the errors in both $n$ and $\Delta P$ increase, as discussed before.
A view of the equation of state as a function of $T/T_c$ for several different
$\mub$ is shown in \fgn{tnpress}. Interestingly, the errors are also large
at $T_c$. However, this is due to the large errors visible in the measurements
of the QNS (see \fgn{ord4}), and not to the resummation.

$\chi_B$ is a thermodynamic response function, in that it determines
the response of QCD matter to a change in $\mub$. QCD at finite $\mub$
has another thermodynamic response function, namely the isothermal bulk
compressibility, $\kappa$. See \apx{kappa} for a definition. Our results
for the equation of state allow us to determine $\kappa$. The results
are shown in \fgn{kappa}.

We also investigated the effect of a change in the quark mass. In
\fgn{mpress} we compare $n$ and $\Delta P$ at two different $T/T_c$, for
the two extreme quark masses. The figure shows that the effect of quark
mass is hardly visible in the data.  The scaling direction $H$ could have
been a mixture of the quark mass and the chemical potential, as discussed
in \apx{widom}.  The observation that the quark mass dependence of the
pressure is statistically insignificant supports an interpretation that
the scaling direction is close to $\mu$. As a result the order parameter,
which is the conjugate thermodynamic variable, must be essentially $n$,
with perhaps a small admixture of the chiral order parameter. This would
further imply that $\psi\simeq0.79$, with little correction. It would
be interesting to extend these computations closer to the chiral and
continuum limits to check whether there are systematic changes as one
approaches the QCD tricritical point.

\section{Conclusions}\label{sec:conclude}
We have made a high statistics study of quark number susceptibilities
in QCD with two flavours of light quarks, for three different sets of
quark masses yielding $m_\pi/m_\rho$ between about 0.6 and 0.25. This
included a very high statistics set (set B) with quark mass tuned to
give $m_\pi/m_\rho\simeq0.3$, extending an earlier study \cite{nt4}.
The number of fermion source vectors used in the stochastic evaluation
of traces is 2000 or more, which is 20 times larger than the number used
earlier. For this set, the number of uncorrelated gauge configurations
is also 3--4 times larger than before, being 200 now.  The effect of
such enhancement in statistics on thermal order parameters is shown
in \fgn{stats}.

The increased statistics allows us to increase the precision of our
measurements, and thereby access new physics. The enhanced precision
is very clearly seen even in the second order QNS; while these are
consistent with previous measurements \cite{nt4}, the higher statistics
gives smoother results with smaller errors. The fourth order QNS
are also consistent with previous results, albeit with a lower peak
at $T_c$.  Both are shown in \fgn{compqns}. 

We infer from \fgn{massdep} that the values of $\chi_{40}$ and $\chi_B^4$
at the peak do not scale as a power of $m_\pi/m_\rho$.  One could take
this as an indication of the failure of at least one of two assumptions:
that the QNS close to the physical pion mass are dominated by the singular
part of the free energy in the chiral limit, or that the scaling in
\eqn{fkrscaling} holds. Such a conclusion may be further reinforced by
the observation of the sixth order QNS, shown in \fgn{ord4}, which peaks
at $T_c$ instead of vanishing. We have discussed that this preliminary
conclusion needs to be tested in two ways.  Finite size scaling studies
would provide more accurate tests of \eqn{fkrscaling}. It would also be
interesting to repeat these computations at lower quark masses to see
what happens closer to the chiral limit.

The QNS can be analyzed to find a radius of convergence of the series
expansion of $\chi_{20}$ via a ratio test, and to infer the existence of
a critical end point, as has been done before. We extend this analysis
here, by noting that when the series diverges, then the truncated
expansion cannot give a good estimate of the values of the QNS, $n$,
and $\Delta P$ near the radius of convergence. If the divergence is
due to a critical point, then $\chi_B\simeq(\mu^2-{\mube}^2)^{-\psi}$.
The logarithmic derivative of $\chi_B$, which is called $m_1$, then
has a pole. A Pad\'e analysis of $m_1$, called a DLOG Pad\'e analysis,
can then be used to resum the series.  Error propagation in this process
is non-trivial, and is described in \apx{pade}. This DLOG Pad\'e tests
whether the divergence is due to a critical point in three ways--- first
by testing whether the pole is at real $\mub$, second by testing whether
the singularity in $m_1$ corresponds to a divergent $\chi_B$, and third by
checking whether the series for $m_1$ can indeed be summed into a single
pole. We find all three criteria yield a signal of a critical point at
the positions given in \tbn{cep}. Extrapolations of our results to the
physical value of $m_\pi/m_\rho\simeq0.18$ gives $T^E/T_c\simeq0.95$
and $\mube/T^E \simeq1.5$. Although the errors are currently too large
to extract a critical index, $\psi$, with any degree of precision, it is
possible to check whether the Ising value, $\psi=0.79$ (see \apx{widom})
is consistent with the data. We find that it is, although the data is
not yet able to distinguish between Ising and mean-field behaviour.

Integrating $m_1$ once and exponentiating, one gets $\chi_B$. Integrating
repeatedly, one gets the baryon density, $n$, and the excess pressure,
$\Delta P$. These are shown in \fgn{npress} as functions of $T$
and $\mub$.  The equation of state is shown as a function of $T/T_c$
at several different $\mub$ in \fgn{tnpress}.  From these quantities it
is also possible to extract the isothermal compressibility of QCD, which
is shown in \fgn{kappa}.  In the previous section we have described in
detail the evidence for a power-law singularity in the second derivative
of $\Delta P$, \ie, the derivative of $n$. Our figures for the equation
of state show how difficult it is to pick out such a mild singularity
from a plot of the function.

In \fgn{mpress} we showed that the effect of quark mass on the equation of
state is smaller than the statistical errors. Along with our observation
of the lack of chiral effects in the QNS, this implies that the scaling
directions for the critical scaling function are close to the physical
parameters $T$, $\mu$ and $m_\pi$, at least in the vicinity of the
physical pion mass. Lattice computations of quantities such as this would
be able to constrain effective theories of the QCD critical point.

These computations were carried out on the Cray XK6 of the Indian
Lattice Gauge Theory Initiative in IACS, Kolkata and the Cray X1
in TIFR, Mumbai. SG would like to thank Saumen Datta, Rajiv Gavai and Rishi
Sharma for discussions and the Department of Science and Technology,
Government  of India, for support under grant no. SR/S2/JCB-100/2011.

\vfill\eject
\appendix

\section{Errors in Pad\'e approximants}\label{sec:pade}
We would like to evaluate the $[0,1]$ Pad\'e approximant
\beq
   P(z;a) = \frac1{z-a},
\eeq{pade}
at various $z$ when $a$ is determined from lattice simulations with
some errors.  If $a$ has Gaussian errors, then, no matter which $z$ we
choose, there is a non-vanishing probability that $a=z$. So the mean and
variance of $P$ both diverge.  To see this, scale and shift $z$ so that
$a$ is distributed as a Gaussian with mean 1 and variance $\sigma^2$. Then
the probability distribution of $P$ at any fixed $z$ is given by
\beq
   p(P;z) = \frac1{\sqrt{2\pi\sigma^2}}\,\frac1{P^2}
         {\rm e}^{-(z-1-1/P)^2/(2\sigma^2)}.
\eeq{dispade}
The distribution is normalizable but none of its moments exist, which means
that the values of the function $P(z;a)$ are completely uncertain.

Fortunately, there is a meaningful way to regularize this problem and to
obtain a finite value and error for the Pad\'e approximant. The solution
lies in recognizing that one always deals with finite statistics, so the
maximum and minimum values of the Pad\'e are always bounded. The analysis
also leads us to a better understanding of the bootstrap procedure.

Suppose that the estimates of the expectation and errors in $a$ are made
with finite statistics, $N$. If one estimates the mean and error of $P(z;a)$ by a bootstrap,
then one should take the
number of bootstrap samples to be ${\cal O}(N)$. In the bootstrap
sample $|z-a|$ has a minimum and a maximum value. The probability that
the minimum value is exactly zero is vanishingly small. Then the maximum
value of $P$ is finite. By accounting
for the restricted range $|P|\le\Lambda$, all the integrals are
regularized.  If the measurements are made with statistics of $N$,
then $\sigma^2\propto1/N$.  In most samples of the bootstrap, one can
find a $\Lambda$ such that
\beq
   \epsilon(\Lambda) = 1 - \int_{-\Lambda}^\Lambda dP p(P;z),
\eeq{epsilon}
and $N\epsilon(\Lambda)\ll 1$. If so, then the problem is regularized
for any fixed value of $N$, in the sense that the bootstrap estimation
yields a finite mean and a finite variance for $P(z;a)$.

The next question is whether one can take the limit $N\to\infty$ with
$\sigma^2\propto1/N$ and $\epsilon\propto1/N$ in such as way that
the mean and variance of $P$ remain bounded. To check this we note that
with increasing $N$ one can arrange $N\epsilon$ to be constant by scaling
$\Lambda\to\zeta\Lambda$ with $\zeta\propto N^{3/2}$. For Gaussian
distributed $a$, the change in the mean and variance when $\zeta$ changes
to $\zeta'$ is
\beqa
\nonumber
   \delta\langle P\rangle &\simeq& {\rm e}^{-K(1-z)^2N} \log(\zeta/\zeta') \\
   \delta\langle P^2\rangle &\simeq& {\rm e}^{-K(1-z)^2N} (\zeta-\zeta'){\Lambda\sigma}
\eeqa{limit}
As a result a bootstrap estimation will lead to bounded mean and error
for the Pad\'e approximant except when $|z-1|<{\cal O}(1/\sqrt N)$.

One can go beyond the Gaussian approximation for the distribution of $a$.
The main idea is to bound the growth of $\langle P\rangle$ and $\langle
P^2\rangle$ by verifying that the estimate of the error in the pole
narrows with $N$ faster than the growth of the probability in the tail
of the distribution of the value of $P(z;a)$.

\section{Widom scaling ansatz}\label{sec:widom}
Consider a system in which the order parameter, $M$, has a non-zero value
only on one side of a temperature
$T=T_c$. Widom scaling is the statement that the ordering field, $H$,
the reduced temperature, $t=T/T_c-1$, and $M$ are related through a
homogeneous function, of degree $\beta\delta$, which can be written in the form
\beq
   H = |M|^\delta J\left(\frac{|t|}{|M|^{1/\beta}}\right).
\eeq{widom}
The exponents $\beta$ and $\delta$, and the function $J$ define the
universality class. Two simple observations follow.  First, if, the system
has finite $M$ for $H=0$, for some $t$, then $|t|/|M|^{1/\beta}$ must be
fixed to be the value which gives $J=0$. Then one has the scaling relation
\beq
   |M| \propto |t|^\beta.
\eeq{scaling}
The second observation is that one can construct the order parameter
susceptibility, $\chi$, by taking the derivative
\beq
   \chi^{-1} = \left.\frac{\partial H}{\partial M}\right|_t =
      |M|^\delta\;\frac{J'}\beta\;\frac{|t|}{|M|^{1+1/\beta}}.
\eeq{grunge}
This is obtained using the chain rule and dropping the term in which
the derivative lands on the prefactor of $J$, since $J=0$. $J'$ is the
value of the derivative for the argument which gives $J=0$. This immediately
gives the Widom scaling formula
\beq
   \chi\propto|t|^{-\gamma},\qquad{\rm with}\qquad \gamma=\beta(\delta-1).
\eeq{scaling2}
These scaling relations are well known.

Applying this to QCD, one would naturally want to interpret
$M=n-n_{\scriptscriptstyle E}$, \ie, the departure of the baryon density,
$n$, from its value at the critical point.  One could set $t$ to be
$\Delta T$, the difference between the temperature and its critical
value in QCD and $H$ to be, $\Delta\mub$, the difference between the
chemical potential and its critical value in QCD. With this simple
identification, one finds the QCD interpretation of the above critical
exponents. However, in QCD, it is more interesting to examine small
$|t|$, \ie, the neighbourhood of the critical point. On doing this,
one finds
\beq
   M\propto H^{1/\delta}, \qquad{\rm and}\qquad \chi\propto H^{-\psi}
    \qquad{\rm with}\qquad\psi=1-\frac1\delta.
\eeq{altscale}
The continuity of pressure forces $\psi$ to be less than unity.
In the 3d Ising model, $\beta=0.33$ and $\delta=4.79$, so one has
$\gamma=1.24$ and $\psi=0.79$. In the mean-field model, $\delta=3$,
which implies that $\psi=0.66$.

\bef
\includegraphics{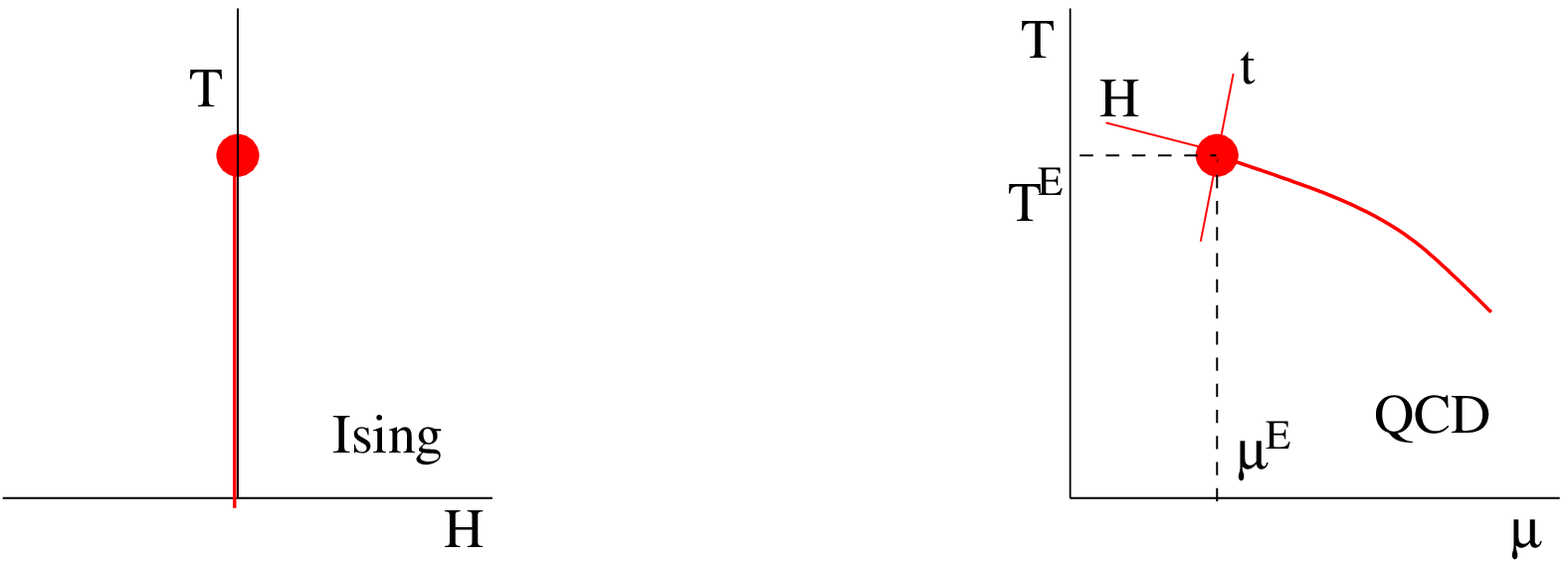}
\caption{The scaling directions $t$ and $H$ in the Ising model are parallel
 to the directions of physical temperature and magnetic field. In QCD,
 however, they may not coincide with $T$ and $\mub$ because the phase
 boundary is curved.}
\eef{critrot}

In QCD there is no compelling reason to declare that the scaling
directions $t$ and $H$ are coincident with the thermodynamic couplings
$\Delta T$ and $\Delta\mub$. The most general course of action
would be to set $t(\Delta T,\Delta\mub)$ and $H(\Delta T,\Delta\mub)$.
When the arguments are small, the functions can be treated in a linear
approximation, which corresponds to scaling (which is absorbed into a
choice of units) and a rotation. So, in general $t = \cos\phi\Delta T +
\sin\phi\Delta\mub$ and $H = \cos\phi\Delta\mub - \sin\phi\Delta T$. The
results of \eqn{altscale} we obtained by taking $\phi=0$.  Instead, if
$\phi=\pi/2$, then $t=\Delta\mub$ and $H=-\Delta T$, and the scaling
formula becomes
\beq
   |\Delta T| = |n|^\delta J\left(\frac{|\Delta\mub|}{|n|^{1/\beta}}\right).
\eeq{wisdom}
Now the analogue of \eqn{scaling} is $n\propto|\Delta \mub|^\beta$.
The scaling of $\chi$
with $\mub$ can be obtained by taking the variation of \eqn{wisdom}
with fixed $\Delta T$ while varying $n$ and $\Delta\mub$ simultaneously:
\beq
   d|\Delta T| = \delta |n|^{\delta-1} J dn + |n|^\delta J'
    \frac{d\Delta\mub}{|n|^{1/\beta}} - |n|^\delta J' \frac{\Delta\mub dn}{|n|^{1+1/\beta}}.
\eeq{grungea}
Taking this at $J=0$, so that the above scaling holds, we find
$\psi=1+\beta=1.33$. The fact that the pressure must be continuous and
finite across the critical point implies that $\psi\le1$. This rules
out $\phi=\pi/2$.  However, by varying the angle $\phi$ one can exhaust
the range $1-1/\delta\le\psi\le1$. The upper limit constrains the scaling
direction $H$ to be close to $\mub$.

More realistically, one should consider the phase diagram in the
space of $T$, $\mub$ and the quark mass. In this extended space, the
scaling directions may be rotated as above, leading to a mixing between
the conjugate variables, the chiral condensate and the baryon density
\cite{dynamics}. As above, this could also lead to an apparent departure
from Ising value, $\psi=0.79$. Then the number density would depend on
$m_\pi$ as the power $2/\delta$. In the Ising model $2/\delta=0.42$, and
in the mean field theory it is 0.66, so the dependence would be
fairly strong.

\section{The isothermal bulk compressibility}\label{sec:kappa}
For materials whose constituents move non-relativistically, the bulk
compressibility, $\kappa$, is defined as
\beq
   1/\kappa = -V \frac{\partial P}{\partial V},
\eeq{comp}
where the other thermodynamic quantities are held fixed as the
pressure and volume are varied. Which quantities are held fixed
defines the ensemble which should be used when computing it from the
underlying theory.  The most commonly used parameter is the isothermal
bulk compressibility, which is defined at fixed particle number and
temperature.

In a relativistic theory the particle number is not conserved, so the
extension to QCD, especially at high-temperatures, requires the usual
generalization:  the canonical ensemble consists of states with fixed
flavour quantum numbers, such as the baryon number, $B$. Thus, one may
generalize the isothermal bulk compressibility to QCD matter by the
definition
\beq
   1/\kappa = -V \left.\frac{\partial P}{\partial V}\right|_{BT}.
\eeq{iso}
The pure gauge theory has infinite $\kappa$; when a gluon gas is
compressed at fixed $T$, the pressure does not change.  Defining $\mub$
by the relation between the internal energy and the baryon number,
$dU=-\mub dB$, when keeping everything else fixed, one may write
\beq
   1/\kappa = V \frac{\partial P}{\partial\mub}
      \,\frac{\partial\mub}{\partial V} = n\mub,
\eeq{canon}
where $n$ is the baryon density. We use this equation to determine
$1/\kappa$. When $\mub$ and $n$ are small then one can make the
approximation $n=\chi_B\mu_B$, and  $\kappa n^2=\chi_B$. It is clear
that this approximation fails as soon as the contribution of higher
order QNS become appreciable. While this is not apparent from \fgn{kappa},
it is clear if one plots $\kappa\mub^2$ against $\mub$.

\end{document}